\documentclass[journal]{IEEEtran}
\usepackage{titlesec}
\titlespacing*{\section}{0pt}{*1}{*1}
\titlespacing{\subsection}{0pt}{*1}{*1}
\usepackage[dvipsnames]{xcolor}
\renewcommand{\thesubsubsection}{\arabic{subsubsection}}
\usepackage{verbatim}
\usepackage{amssymb,amsfonts}
\usepackage{amsthm}

\usepackage[linesnumbered,ruled,vlined]{algorithm2e}

\usepackage{array}
\usepackage[caption=false,font=normalsize,labelfont=sf,textfont=sf]{subfig}
\usepackage{textcomp}
\usepackage{stfloats}
\usepackage[T1]{fontenc}
\usepackage{balance}
\usepackage{enumitem}
\usepackage{url}
\usepackage{cuted}
\usepackage[makeroom]{cancel}
\usepackage{lipsum}
\usepackage{verbatim}
\usepackage{graphicx}
\usepackage{cite}
\usepackage{amsmath,amssymb,amsfonts}
\usepackage{mathtools}
\usepackage{graphicx}
\usepackage{textcomp}
\usepackage{xcolor}
\usepackage{listings}
\usepackage{pdfpages}
\usepackage{mathrsfs}
\usepackage{bm}
\usepackage[caption=false]{subfig}
\usepackage{amsthm}
\theoremstyle{definition}
\newtheorem{theorem}{Theorem}
\newtheorem{lemma}{Lemma}
\newtheorem{proposition}{Proposition}
\newtheorem{remark}{Remark}

\newtheorem{corollary}{Corollary}
\usepackage{mathrsfs}
\def\BibTeX{{\rm B\kern-.05emb{\sc i\kern-.025em b}\kern-.08em
    T\kern-.1667em\lower.7ex\hbox{E}\kern-.125emX}}

\usepackage{tikz}
\usepackage{circuitikz}
\usetikzlibrary{positioning}
\usepackage{pgfkeys}
\usetikzlibrary{spy}

\usepackage{pgfplots}

\usepgfplotslibrary{groupplots} 
\usepgfplotslibrary[groupplots] 
\usetikzlibrary{pgfplots.groupplots} 
\usetikzlibrary[pgfplots.groupplots] 

\pgfplotsset{compat=newest}
\pgfplotsset{plot coordinates/math parser=false}
\usepackage{pgfplotstable}
\newlength\figureheight
\newlength\figurewidth

\usetikzlibrary{positioning}
\usetikzlibrary{patterns}

\usetikzlibrary{arrows.meta,positioning}
\tikzset{block/.style={draw, rectangle, fill=cyan!90,
        minimum height=2em, minimum width=3em},
    sum/.style={draw, circle, node distance=1cm},
    input/.style={coordinate},
    output/.style={coordinate},
    pinstyle/.style={pin edge={to-,thin,black}},
        saturation block/.style={%
            draw,
            path picture={
                \pgfpointdiff{\pgfpointanchor{path picture bounding box}{south west}}%
                {\pgfpointanchor{path picture bounding box}{north east}}
                \pgfgetlastxy\x\y
                \tikzset{x=\x*.4, y=\y*.4}
                \draw [very thin] (-1,0) -- (1,0) (0,-1) -- (0,1);
                \draw [very thick] (-1,-.7) -- (-.7,-.7) -- (.7,.7) -- (1,.7);
            },
        }
    }

\tikzset{%
        rateLimit block/.style={%
            draw,
            path picture={
                \pgfpointdiff{\pgfpointanchor{path picture bounding box}{south west}}%
                {\pgfpointanchor{path picture bounding box}{north east}}
                \pgfgetlastxy\x\y
                \tikzset{x=\x*.4, y=\y*.4}
                \draw [very thin] (-1,0) -- (1,0) (0,-1) -- (0,1);
                \draw [very thick] (-1,-1) -- (1, 1);
            },
        }
    }

     \usetikzlibrary{svg.path}
     \usepackage{scalerel}
\usepackage{hyperref}
    \definecolor{orcidlogocol}{HTML}{A6CE39}
    \tikzset{
      orcidlogo/.pic={
        \fill[orcidlogocol] svg{M256,128c0,70.7-57.3,128-128,128C57.3,256,0,198.7,0,128C0,57.3,57.3,0,128,0C198.7,0,256,57.3,256,128z};
        \fill[white] svg{M86.3,186.2H70.9V79.1h15.4v48.4V186.2z}
                     svg{M108.9,79.1h41.6c39.6,0,57,28.3,57,53.6c0,27.5-21.5,53.6-56.8,53.6h-41.8V79.1z M124.3,172.4h24.5c34.9,0,42.9-26.5,42.9-39.7c0-21.5-13.7-39.7-43.7-39.7h-23.7V172.4z}
                     svg{M88.7,56.8c0,5.5-4.5,10.1-10.1,10.1c-5.6,0-10.1-4.6-10.1-10.1c0-5.6,4.5-10.1,10.1-10.1C84.2,46.7,88.7,51.3,88.7,56.8z};
      }
    }

    \newcommand\orcidicon[1]{\href{https://orcid.org/#1}{\mbox{\scalerel*{
    \begin{tikzpicture}[yscale=-1,transform shape]
    \pic{orcidlogo};
    \end{tikzpicture}
    }{|}}}}

\begin{document}



\title{Enhanced BICM Receivers for Ultra-Reliable\\Low-Latency Short Packet Communications}

\author{\IEEEauthorblockN{ Mody~Sy~{\orcidicon{0000-0003-2841-2181}}},~\IEEEmembership{Member IEEE}~and~ Raymond~Knopp~{\orcidicon{0000-0002-6133-5651}},~\IEEEmembership{Fellow IEEE}

\thanks{Manuscript prepared on October 06, 2025. The authors are with the Department of Communication Systems, EURECOM, BIOT, 06410, France (e-mails: mody.sy@eurecom.fr).}

\thanks{This article was in part presented at the 2023 IEEE Conference on Standards for Communications and Networking (Munich, Germany, November 2023) and the 2024 European Conference on Networks and Communications \& 6G Summit (Antwerp, Belgium, June 2024).}
}

\makeatletter
\patchcmd{\@maketitle}
  {\addvspace{0.5\baselineskip}\egroup}
  {\addvspace{-1\baselineskip}\egroup}
  {}
  {}
\makeatother
\maketitle
\begin{abstract}
This paper presents enhanced receiver metrics for joint estimation-detection in short blocklength transmissions, addressing scenarios with unknown channel state information and low  or sparse training resource density. We show that it is possible to enhance the performance and sensitivity through block-wise joint estimation–detection compared to standard receivers. The performance analysis makes use of a full 5G transmitter and receiver chains for both Polar and LDPC coded transmissions paired with QPSK modulation scheme. We consider transmissions where reference signals are interleaved with coded data and both are transmitted over a small number of OFDM symbols so that near-perfect channel estimation cannot be achieved.    Unlike conventional symbol-by-symbol detection in BICM systems, where the observation for a given coded bit is confined to the symbol in which it is conveyed,the proposed method performs block-wise joint detection over a sliding window of adjacent symbols to  fundamentally leverages their statistical dependencies. Accordingly, the LLR for a particular coded bit incorporates information from all symbols within the detection window, rather than being constrained to its host symbol alone. Performance evaluation spans SIMO and SU-MIMO configurations, emphasizing the efficacy of the estimation-detection strategy in realistic base station receiver scenarios. Our findings demonstrate that when the detection windows used in the metric units are on the order of four modulated symbols, the proposed receivers remarkably outperform the conventional ones and can be used to achieve detection performance that is close to that of coherent receivers with perfect CSI.
\end{abstract}

\section{Introduction}

\IEEEPARstart{I}{t is expected} that the 6G air-interface will build upon the 5G standard and address new paradigms for feedback-based cyber-physical systems combining
communications and sensing. In particular, the deployment of 6G-enabled devices will necessitate ultra-tight control loops over the air interface, demanding unprecedented levels of reliability and latency potentially exceeding the sub-millisecond uplink application-layer latency targets in the microwave spectrum currently set by 5G.  This critical requirement for ultra-reliable and low-latency communications fundamentally drives ongoing advancements in the physical layer, with a particular emphasis on the design of highly sophisticated receivers. Notably, reliability emerges as a cornerstone metric, underpinning receiver architectures designed to satisfy the stringent key performance indicators (KPIs) imposed by beyond-5G (B5G) and 6G wireless communication standards.

However, although 5G transmission formats can provide very short-packet transmission through the use of mini-slots, the ratio of training information to data is not necessarily adapted to extremely short data transmission. Moreover, these formats are primarily designed under the assumption of conventional quasi-coherent receivers, which can become highly suboptimal in scenarios where accurate channel estimation is infeasible either due to the sporadic nature of 
of such transmissions or to stringent constraints on decoding latency.

This challenge is particularly emerging in so-called {\em ultra-reliable-low-latency communication} (URLLC) industrial IoT applications.
This would be similar for evolved channel state information (CSI) feedback control channels, or for future joint-sensing and communication paradigms requiring rapid sensory feedback to the network. One of the main applications is in the field of mission-critical communications, such as those used by emergency services or in industrial control systems that require extremely high levels of reliability and low latency. In this work, we investigate {\em bit-interleaved coded modulation} receiver design for short data transmission. Specifically, we are interested in designing joint estimation–detection based receivers compliant with polar and {\em low-density parity-check} (LDPC) coded modulation transmission, targeting short packets in the range of 20–100 bits for the envisaged beyond 5G/6G signaling scenarios. Indeed, BICM remains a widely adopted coded-modulation technique for error-prone wireless channels. On the receiver side, its efficacy hinges on the underlying detection and decoding metrics, underscoring the trade-off between enhanced performance and low complexity. Noteworthy is the historical integration of BICM into 3GPP systems, a practice dating back to the 3G-era.

Furthermore, there is a wealth of literature on BICM receivers from various perspectives \cite{Fabregas08}\cite{CTB98} demonstrating their potential impact and importance. Among the pioneers who sparked interest in BICM was the seminal work conducted by Caire \textit{et al.} \cite{CTB98}, wherein they provided a comprehensive analysis in terms of information rate and error probability. Afterwards, numerous research inquiries have been directed towards the design of reliable low-complexity receivers for MIMO ({\em multiple input multiple output}) and Non-MIMO BICM systems, but mostly restricted to coherent communication scenarios. Staring closely at the primary focus of this investigation, namely the transmission of short packets, it is evident that this area has garnered significant scholarly attention in recent years. Extensive research has addressed various aspects, including the design of signal codes and receiver algorithms \cite{Lee2018, sy2023_1, sy2023_2,  sy2023_3, Doan2022, Yue2023, Gamage2020, Zheng2024, Mestre2020, Cao2022}, as well as the establishment of state-of-the-art converse and achievability bounds \cite{Xhemrishi2019, Polyanskiy2010, Durisi2016, Ostman2019jrnal, Martinez2011}.

This work distinguishes itself from prior literature by introducing novel enhanced receiver designs tailored to scenarios with imperfect {\em channel state information} (CSI) over various channel conditions. As previously mentioned, the proposed receivers are specifically designed to effectively support the reception of short data packets in beyond 5G/6G signaling scenarios by evaluating their performance over 5G short block channels using both Polar and LDPC coded modulation formats.
(a) We look into receiver metrics exploiting {\em joint estimation and detection} (JED), which are particularly amenable to configurations where low-density of {\em demodulation reference signals} (DMRS) are interleaved with coded data symbols.
(b) We specifically address situations where accurate channel estimation is impossible, demonstrating that a well-conceived joint estimation–detection receiver, leveraging interleaved DMRS within the detection metric, can achieve remarkably significant performance gains over conventional 5G {\em orthogonal frequency division multiplexing} (OFDM) receivers, and can potentially approach the performance of a perfect CSI receiver, applicable to both uplink and downlink transmission scenarios.

Explicitly, our proposal consists of designing and utilizing novel soft-likelihood metrics that directly integrate channel estimation performed via joint least squares followed by averaging or smoothing across DMRS dimensions for bit-level LLR generation. Moreover, we apply to the underlying soft-likelihood metrics an advanced block-wise joint detection scheme defined over a detection window of four modulated symbols ($M=4$). Unlike conventional symbol-by-symbol detection in BICM systems, where the observation for a given coded bit is confined to the symbol in which it is conveyed, the proposed block-wise detection approach fundamentally leverages the statistical dependencies between adjacent symbols; that is, the LLR for a given coded bit incorporates information from all symbols within the detection window, rather than being constrained to its host symbol alone, thereby enhancing detection reliability. Hence, our contributions span the following principal avenues. Initially, we introduce a  BICM receiver metric specifically tailored for non-coherent fading channels in  {\em single-input multiple-output} (SIMO) transmissions. These metrics effectively address challenges arising from both line-of-sight (LOS) and {\em non-line-of-sight} (NLOS) fading channels. Secondly, we extend the BICM receiver metric design  to {\em single-user} MIMO systems, specifically addressing block fading channel conditions.

The article  is structured as follows. Section II lays out the system model and the foundations of 5G polar and LDPC coded modulations, Section III focuses on the proposed BICM receiver metrics, Section IV presents the numerical results and performance analysis, and finally Section V concludes the paper.

{\em Notation :}
Scalars are denoted by italic letters, vectors and matrices are denoted by bold-face lower-case and upper-case letters, respectively.
 For a complex-valued vector $\mathbf x$, $|\lvert \mathbf x |\rvert$ denotes its Euclidean norm, $| \cdot |$  denotes the absolute value.  $\| \cdot \|_\mathsf{F}$ is the Frobenius norm of matrix.  $ \operatorname{tr}\{\cdot\}$ denotes the trace of matrix.
   $ \mathbb E\{\cdot\}$ denotes the statistical expectation. $\operatorname{Re}(\cdot)$ denotes the real part of a complex number. $\operatorname{I_0}(\cdot)$ is the zero-th order modified Bessel function of the first kind.
$\mathbf I$ is  an identity matrix with appropriate dimensions.
 $\mathbf x \in \boldsymbol{\chi}^j_b=\left\{ \mathbf x: e_j=b\right\}$ is the subset of symbols $\mathbf x$ for which the $j-th$  bit of the label $e$ is  equal to $b=\{0,1\}$, $\boldsymbol{\chi}$ is the modulation alphabet (e.g., QPSK, 16-QAM, $\ldots$).  The number of bits required to a symbol is denoted by $m\coloneqq \log_2|\boldsymbol{\chi}|$,  where $|\boldsymbol{\chi}|$ is the  cardinality of $\boldsymbol{\chi}$.
$ \Lambda^j\left(\cdot\right)$ denotes log likelihood ratio, with $j=1,2, \ldots, m$. $M$ is the number of symbols making up a block on which joint detection is performed.
The superscripts  $^\top$  and  $^\dag$ denote the transpose and  the complex conjugate transpose or Hermitian. The operator \(\biguplus\) denotes a \textit{disjoint union}: \(\mathsf{d} = \biguplus_{b=1}^B \mathcal{D}_b\) means that the index sets \(\mathcal{D}_b\) are pairwise disjoint and their union exactly covers \(\mathsf{d}\), i.e.,\(\mathcal{D}_b \cap \mathcal{D}_{b'} = \emptyset \quad \text{for all } b \neq b', \quad \text{and} \quad \bigcup_{b=1}^B \mathcal{D}_b = \mathsf{d}.\)

\section{General Framework}
\subsection{ Bit-Interleaved Polar-coded Modulation (BIPCM)}
 Bit interleaved polar coded modulation is referred to as BIPCM. In this instance, we are dealing with the {\em cyclic redundancy check} (CRC)-aided polar coding scheme, one of the basic code construction techniques established by the 3GPP Standard\cite{3GPP38212}. Using polar codes as a channel coding scheme for 5G control channels has demonstrated the significance of Arikan's invention \cite{Arikan2009}, and its applicability in commercial systems has been  proven. This new coding family achieves capacity rather than merely approaching it as it is based on the idea of channel polarization. Polar codes can be used for any code rate and for any code length shorter than the maximum code length due to their adaptability.
In 5G new radio, the polar codes are employed to encode {\em broadcast channel} (BCH) as well as {\em downlink control information} (DCI) and {\em uplink control information} (UCI). Furthermore, the transmission process is straightforward and complies with the 3GPP standard specifications \cite{3GPP38212}.

With respect to the decoding process, several main polar code decoding algorithms are currently used, including the SC algorithm \cite{Arikan2009}, the {\em successive cancellation list} (SCL) algorithm \cite{Stimming2014}\cite{Tal_vardy2015}, the {\em CRC-aided SCL} (CA-SCL) algorithm \cite{Zhang2017}\cite{Niu2012}, the {\em belief propagation} (BP) algorithm \cite{Arikan2008}, and the {\em successive cancellation with adaptive node} (SCAN) algorithm \cite{Fayyaz2013}. 
The SCL algorithm improves upon the SC algorithm by providing multiple paths and outperforms it in terms of performance. The CA-SCL algorithm incorporates a high-rate CRC code to assist in selecting the correct codeword from the final list of paths in the SCL decoder, effectively enhancing its reliability. It has been observed that the right codeword is usually included in the list every time the SCL decoder fails. \\
The performance ranking of the decoding algorithms is as follows: \texttt{CA-SCL$ > $ state-of-the-art SCL$>$BP=SCAN$>$SC}. Therefore, for improved performance, the channel decoder technique should utilize CA-SCL decoding for downlink (DCI or BCH) or uplink (UCI) messages. The adoption of polar codes by 3GPP was partly due to the well-acknowledged potential of CA-SCL decoding to outperform Turbo or LDPC codes.
%
\subsection{Bit-Interleaved LDPC-coded Modulation (BILCM)}
Bit-Interleaved LDPC-Coded Modulation is referred to as BILCM. First proposed by Gallager in the early 1960s \cite{Gallager63}, LDPC coding has proven to be highly suitable for 5G NR due to its advantages such as high throughput, low latency, efficient decoding complexity, and rate compatibility. The performance of LDPC codes in 5G NR is impressive, exhibiting an error floor at or below the $10^{-5}$ {\em block error rate} (BLER), a significant improvement over traditional coding techniques.
Furthermore, the BILCM transmission procedure is almost identical to that described with BIPCM.


A code block is encoded by the 5G LDPC encoder according to the procedure defined in the 3GPP standard \cite{3GPP38212}.
At the receiver, the LDPC decoding is performed on each code block individually. For LDPC decoding, various techniques can be implemented, with belief propagation (BP) methods being the most commonly used. BP methods rely on iterative message exchange between bit nodes and check nodes, offering near-optimal decoding performance at the cost of computational complexity. However, to strike a better balance between performance and complexity, several simplified and effective decoding algorithms have been proposed in the scientific literature. One such decoding algorithm is layered message passing \cite{Hocevar2004}, which stands out as a promising approach for URLLC due to its ability to speed up convergence times \cite{Zhang2014, wang2021}, making it a suitable candidate for short packet transmissions.\\
The foundations of polar and LDPC coding and decoding are beyond the scope of this paper, but interested readers may wish to refer to one of our prior correspondences \cite{Sy2025}.

\subsection{Modulation and Resource Mapping}
In both scenarios, the encoded payload undergoes rate-matching and code block concatenation prior to being fed into a QPSK modulator. This process yields a set of complex-valued modulation symbols. Subsequently, the resource mapping process is executed, where one or multiple OFDM symbols are used to allocate the modulated symbols to resource blocks and insert the DMRS resources. The number of resource blocks is determined by the payload size and coding settings. When the payload size is small, fewer resource blocks are required, thereby maintaining a constant effective coding rate.

Furthermore, the transmitted signal $\mathbf{x}$ typically consists of data-dependent $\mathbf{x}^{(\mathsf d)}$ and data-independent $\mathbf{x}^{(\mathsf p)}$ components, known as pilot or reference signals. The reference signals are used to resolve channel ambiguity across time, frequency, and/or spatial domains.  Specifically, they are employed to estimate the channel.
In practice, the reference  signals are commonly interleaved among the data-dependent components. It is notably the case in current OFDM systems. In earlier CDMA systems, reference  signals were sometimes superimposed on top of data-dependent signals.
The number of data dimensions is denoted by $N_d$, and the number of reference signal dimensions is denoted by $N_p$, where $N_d+N_p=N$.
In 3GPP standard, $N$ is typically equal to $12{K}{L}$. This represents the number of complex dimensions or resource elements(REs) in the {physical resource blocks} (PRBs). The number of PRBs, ${K}$ ranges from $1$ to $16$, while the number of OFDM symbols, ${L}$, ranges from $1$ to $14$, and can be increased if multiple slots are used for signaling the channel bits. Resource elements $\mathbf x =\{\mathrm x_n : n=1,2, \ldots, N\}$ are mapped onto $N$ subcarriers such that $\forall n \in \mathcal N_p \cup \mathcal N_d$ with $\mathcal N_p$ the set of subcarriers for DMRS and $\mathcal N_d$ the set of subcarriers for data. The assumption in this work is that the data-dependent components $\mathbf{x}^{(\mathsf{d})}$ are generated from a binary code whose output is interleaved and subsequently mapped onto an $\mathcal{M}$-ary modulation symbol alphabet. We will assume that the binary code generates $E$ bits and the interleaver mapping is one-to-one so that $E$ bits are also fed to the modulator. The binary-code and interleaver combination can thus be seen as a $(E,B)$ binary block code. We denote the $E$ interleaved coded bits as $e_k,k=0,1,\cdots, E-1$.
Adjacent $\log_2 |\boldsymbol{\chi}|$ bit-tuples are used to select the modulated symbols in the symbol alphabet. Unless otherwise stated, we assume that Gray mapping is applied when using non-binary modulation schemes.
In instance of the quadrature phase-shift keying (QPSK) modulation, pairs of bits $e[2n]$ and $e[2n +1]$, are mapped to complex-valued modulation symbols as follows:
\begin{equation}
\resizebox{0.49\textwidth}{!}{$
\mathrm{x}^{(\mathsf{d})}[n] =
\begin{cases}
\frac{1}{\sqrt{2}}(1 + i),   & \text{if } (e[2n], e[2n+1]) = (0, 0) \\
\frac{1}{\sqrt{2}}(-1 + i),  & \text{if } (e[2n], e[2n+1]) = (0, 1) \\
\frac{1}{\sqrt{2}}(-1 - i),  & \text{if } (e[2n], e[2n+1]) = (1, 1) \\
\frac{1}{\sqrt{2}}(1 - i),   & \text{if } (e[2n], e[2n+1]) = (1, 0)
\end{cases}
\quad \forall n \in \mathcal{N}_d .
$}
\end{equation}

Likewise, it is worth examining how DMRS sequences making up  $\mathbf{x}^{(\mathsf p)}$ are constructed. These sequences are derived from Zadoff-Chu (ZC) sequences, leveraging their desirable properties in terms of ideal auto-correlation, low PAPR (Peak-to-Average Power Ratio), constant amplitude, and their proven standardization adoption in LTE and 5G NR.
In that respect, the DMRS sequence is defined as:
\begin{equation}
\resizebox{0.43\textwidth}{!}{$
\mathrm{x}^{(\mathsf p)}[n] = \begin{cases}
    \exp\left( - i \pi \cdot \frac{u n^2}{N_p} \right) & \text{if }  N_p \text{ even}  \\
    \exp\left( - i\pi \cdot \frac{u n(n+1)}{N_p} \right)& \text{if }  N_p \text{ odd}
\end{cases}, \\ \forall n \in \mathcal N_p,
$}
\end{equation}
where $u$ is the root index, which must be coprime with $N_p$ (i.e., $\texttt{gcd}(u,N_p)=1$). The power of each DMRS sequence is typically normalized to unity.
Furthermore, in spatially multiplexed MIMO systems, the resource mapping procedure must account for pilot symbol allocation not only in time and frequency but also across the spatial dimension. Specifically, training symbols must be transmitted in a way that avoids interference to ensure accurate CSI estimation. This requires careful consideration of pilot allocation strategies, which may exploit frequency orthogonality, time orthogonality, or signal orthogonality. In this study, signal orthogonality is adopted for the MIMO configuration. Indeed, in a MIMO transmission system with ($\mathsf{N_T}\geq 1$), these antennas often share the same time-frequency resources for DMRS. To prevent inter-antenna interference, Zadoff-Chu sequences with the same root are orthogonalized across antennas using cyclic shifts (i.e., linear phase rotations).
Thus we define the DMRS sequence $\mathrm{x}^{(\mathsf p)}_t[n]$ transmitted via antenna port $t=1, 2, \ldots, \mathsf{N_T}$ by
\begin{equation}
\resizebox{0.43\textwidth}{!}{$
\mathrm{x}^{(\mathsf p)}_t[n] = \mathrm{x}^{(\mathsf p)}[n]\cdot \exp\left( i \cdot \frac{2\pi (t-1) n}{\mathsf{N_T}} \right), \
 \forall n \in \mathcal N_p .
 $}
\end{equation}

The second term $e^{i \cdot \phi_t(n)}$, where $\phi_t(n)=\frac{2 \pi(t-1) n}{\mathsf{N_T}}$,  applies a linear cyclic phase shift to orthogonalize the DMRS sequences across different transmit antennas. To ensure this orthogonality, as required by the 5G standard, an additional condition must be satisfied. Indeed, the number of DMRS positions $N_p$ must be at least equal to the number of antenna ports $\mathsf{N_T}$, which means: $\mathsf{N_T}\leq N_p$.\\

\begin{figure*}[ht]\centering
  \includegraphics[width=0.8\linewidth]{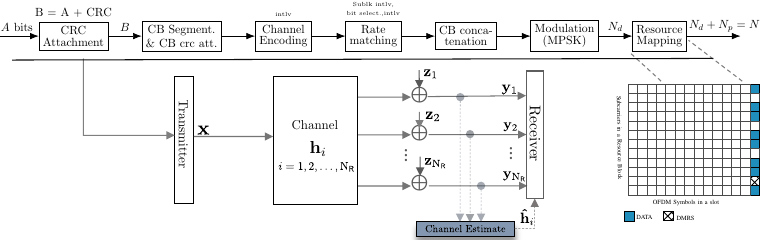}
      \caption{Bit-Interleaved Polar/LDPC coded Modulation ( SIMO BIPCM/ SIMO BILCM) : Transmitter end.}
      \label{fig:bicm_polar_ldpc}
      \vspace{-5mm}
\end{figure*}

Figure~\ref{fig:bicm_polar_ldpc} presents an overview of the BIPCM/BILCM process short block uplink channels. The transmit-end procedure includes several steps such as adding a transport block CRC, segmenting code blocks with additional CRC attachment, channel encoding, rate matching, code block concatenation, and modulation. It is important to emphasize that the receiving chain simply follows the reverse flow of the transmitter-end.

\section{Receiver Design}
\subsection{\texorpdfstring{$(1 \times \mathsf{N_R} )$} \ ~SIMO  Non-Coherent Fading Channel}
Considering a SIMO OFDM BICM system with a single antenna element on the transmit array ($\mathsf{N_T}=1$) and multiple element receive arrays ($\mathsf{N_R}$). The transmitted and received signals are $N$-dimensional column vectors, and thus a system is designed in such a way that the relationship between the transmitted and received signals is as follows:
\begin{equation}
  \mathbf{y}_r = \mathbf h_r\mathbf{x}+ \mathbf{z}_r, \quad r=0,1,\cdots,\mathsf{N_R}-1,
\end{equation}
where $\mathbf z_r$ is additive white Gaussian noise whose  real and imaginary components are independent and have variance and $\mathbf{h}_r$ represents the channel vector.
\subsubsection{Perfect Channel State Information}
 In the instance of perfect channel state information, 
 the likelihood function  is shown to be:
 \begin{equation}\label{eqn:lfpcsi1}
     \resizebox{0.43\textwidth}{!}{$
 \begin{gathered}
   \mathrm q\left(\mathbf{x},\left\{\mathbf{y}_r, \mathbf h_r\right\}\right)=\mathrm p\left(\mathbf{y}_r, \mathbf h_r \mid \mathbf{x}\right)=
   \mathrm p\left(\mathbf{y}_r \mid \mathbf{x}, \mathbf h_r\right) \mathrm p\left(\mathbf h_r \mid \mathbf{x}\right).
   \end{gathered}
   $}
   \end{equation}
 If the transmitted signal ${\mathbf x}$ is independent of the channel realization $\mathbf h_r$, the term $p\left({\mathbf h}_r\mid\mathbf{x}\right)$ in (\ref{eqn:lfpcsi1}) can be dropped since it will disappear in (\ref{eqn:lfpcsi}). The likelihood function is commonly equivalent to :
   \begin{equation}\label{eqn:lfpcsi}
   \begin{gathered}
   \mathrm q\left(\mathbf{x},\left\{\mathbf{y}_r, \mathbf h_r\right\}\right)\propto \exp \left(-\frac{1}{\mathrm {N}_0 }\lvert|\mathbf y_r-\mathbf h_r \mathbf x\rvert|^2\right).
   \end{gathered}
   \end{equation}
Using the norm extension property, ignoring terms that are independent of $\mathbf x$,
the likelihood function then simply becomes:
   \begin{equation}\label{eqn:lfpcsi3}
       \resizebox{0.42\textwidth}{!}{$
   \begin{aligned}
   &\mathrm q\left(\mathbf{x},\left\{\mathbf{y}_r, \mathbf h_r\right\}\right) \propto
   &\exp \left(\frac{ 2}{\mathrm {N}_0}\operatorname{Re}\left(\mathbf{y}_r \mathbf h_r^\dag \mathbf{x}^\dag\right) -\frac{1}{\mathrm {N}_0}\lvert|\mathbf h_r \mathbf{x}\rvert|^2\right).
   \end{aligned}
   $}
   \end{equation}
The likelihood of coded bit $e_j\in\{0,1\}$ is
 \begin{equation}\label{eqn:lf_simo}
   \mathrm q\left(e_j(\mathbf x)=b ,\left\{\mathbf{y}_r, \mathbf h_r\right\}\right)=
   \sum_{\mathbf x \ \in \ \boldsymbol{\chi}_b^j}\mathrm q\left(\mathbf{x}, \left\{\mathbf{y}_r, \mathbf h_r\right\}\right).
   \end{equation}
As is common in the case of BICM-based systems, the soft input to the binary channel decoder is given as the log-likelihood ratio (LLR) for the $j-th$  coded bit, such that :
\begin{equation}\label{eqn:llr_simo}
   \Lambda^j\left(\mathbf y_r\right)=\log \frac{\mathrm q\left(e_j(\mathbf x)=0 ,\left\{\mathbf{y}_r, \mathbf h_r\right\}\right)}{\mathrm q\left(e_j(\mathbf x)=1 ,\left\{\mathbf{y}_r, \mathbf h_r\right\}\right)}.
\end{equation}
We simplify (\ref{eqn:llr_simo}) using a \emph{max-log approximation}: $\log \left\{\sum_{i} \exp \left(\lambda_{i}\right)\right\} \sim \max _{i}\left\{\lambda_{i}\right\}$, resulting in (\ref{eqn:maxlog_simo_conv_rx}).
\begin{equation}\label{eqn:maxlog_simo_conv_rx}
  \resizebox{0.48\textwidth}{!}{$
\begin{aligned}
\Lambda^j\left(\mathbf y\right)&=  \sum_{b\in\{0,1\}} (-1)^{b}\,
   \Biggl[\displaystyle \max _{\mathbf x \ \in \ \boldsymbol{\chi}_b^j}\frac{1}{\mathrm {N}_0}  \left(\sum_{r=0}^{\mathsf{N_R}-1}2\operatorname{Re}\left(\mathbf{y}_r \mathbf h_r^\dag \mathbf{x}^\dag\right) -\lvert|\mathbf h_r \mathbf{x}\rvert|^2\right)  \Biggr].
\end{aligned}
   $}
\end{equation}
This metric is typically used in {\em Perfect CSI} based receivers as well as in conventional quasi-coherent receivers which employs least-squares channel estimation followed  by linear interpolation, thereby replacing $\mathbf{h}$ with $\hat{\mathbf h}$. 
We consider these receivers as a benchmark for comparison with the subsequent proposed receivers.

Moreover, within the framework of a conventional receiver, it is presupposed that, at the very least, the observation of a single reference signal spans a PRB in order to generate the coded bits corresponding to each data symbol in that PRB.
In addition, this conventional metric applies a so-called {\em symbol-by-symbol} detection, that’s to say that each symbol is detected independently of the other symbols.
We will note this as {\em No CSI} Conv.
\subsubsection{Unknown Channel State information}
Non-coherent fading channels refer to communication channels in which the fading coefficients are not known a priori and must be estimated, thereby requiring detection or estimation techniques that do not rely on explicit CSI. In what follows, we describe BICM metrics for a general non-coherent fading channel with unknown phase on the line-of-sight (LOS) components and fully unknown diffuse (Non-LOS) components. The overall unknown channel gain is given by
$\mathbf {h}_r=\left(\sqrt{\alpha}e^{j\theta_r}+\sqrt{1-\alpha}\mathrm h_r^{(f)}\right)\mathbf{I}$, where $\theta_r$ is assumed to
be i.i.d. uniform random variables on $[0,2\pi)$, $\mathrm h_r^{(f)}$ is a zero-mean, unit-variance, circularly-symmetric complex Gaussian random variable and $\alpha$ is the relative strength of the LOS component.
The amplitude $|\mathrm h_r|$ on each receive branch is thus Ricean distributed. It is worth noting that
the i.i.d. assumption for the ${\theta_r}$ is somewhat unrealistic for a modern array receiver with accurate calibration.
The phase differences would be more appropriately characterized by two random-phases, one originating from the time-delay between transmitter and receiver and the other from the angle of arrival of the incoming wave. The phase differences of individual
antenna elements for a given carrier frequency could then be determined from the angle of arrival and the particular
geometry of the array. To avoid assuming a particular array geometry, the i.i.d. uniform model provides a simpler and universal means to derive a receiver metric.

\begin{proposition}{\em [A novel soft-likelihood metric for SIMO.]}\label{prop:4.1}~\\
Neglecting multiplicative terms independent of the transmitted message, the likelihood function can be expressed as follows:
\begin{equation}\label{eqn:llrfunction_simo_non_coh_fad}
\begin{aligned}
\mathrm q\left(\mathbf{x},\mathbf y\right)&= \prod_{r=0}^{\mathsf{N_R}-1} \frac{1}{\mathbf L_{\mathsf x}}\exp \left(  -\frac{ \alpha \left\|\mathbf x \right\|^2 }{\mathbf L_{\mathsf x}}  + \right. \\& \left. \beta_x \left| \mathbf x^\dag \mathbf y_r\right|^2\right)\times\operatorname{I_0}\left(  \frac{2\sqrt{\alpha}}{\mathbf L_{\mathsf x}}\left|\mathbf x^\dag\mathbf y_r\right|\right),
\end{aligned}
\end{equation}
where
$\mathbf L_{\mathsf x} = \mathrm {N}_0+ 2(1-\alpha) \left\|\mathbf x \right\|^2$,  $\mathbf \beta_x = \frac{2(1-\alpha)}{\mathrm {N}_0(\mathrm {N}_0 +2(1-\alpha) \left\|\mathbf x \right\|^2)}$ and $\operatorname{I_0}(\cdot)$ is the zero-order
modified Bessel function.
\end{proposition}
\proof see Appendix Section A. \qedhere\\
We then apply equations (\ref{eqn:lf_simo})–(\ref{eqn:llr_simo}) to succinctly generate the LLR of the $j$-th coded bit.

Note that in the above expressions, we do not limit the dimensionality of the observations when computing likelihoods of particular bits. In the original work of Caire\text{ et al.} \cite{CTB98}, the authors assume an ideal interleaving model which allows limiting the observation interval of a particular coded bit to the symbol in which it is conveyed. For long blocks, this assumption is realistic for arbitrary modulation signal sets and is sufficient for BPSK and QPSK irrespective of the block length when the channel is known perfectly. Nevertheless, practical systems usually apply single symbol likelihood functions for short blocks and high-order modulations. Furthermore, for the primary case of interest here, namely transmission without channel state information, single symbol detection is impossible. At the very least, the observation of one reference symbol must be used to generate likelihoods of the coded bits of a data symbol, thus warranting the study of block detection.\\

Furthermore, the LLR metric calculations based on (\ref{eqn:llrfunction_simo_non_coh_fad}) in the logarithmic domain can be hard to implement and computationally prohibitive. A common simplification is the \emph{max-log approximation}. First, an exponential approximation is applied to the modified Bessel function of the first kind $\operatorname{I_0}(z)$, which results in $\operatorname{I_0}(z) \sim \frac{e^z}{\sqrt{2\pi z}} \sim e^z$.

Thus the maxlog-likelihood ratio for the $j-th$ coded bit is given by (\ref{eqn:maxlog_llr_simo}).
\begin{align}
  \label{eqn:maxlog_llr_simo}
  \resizebox{0.42\textwidth}{!}{$
\Lambda^j(\mathbf y)
= \displaystyle \sum_{b\in\{0,1\}} (-1)^{b}\,
   \Biggl[
     \max_{\mathbf x\in\boldsymbol{\chi}_b^j}\Bigl(
       \sum_{r=0}^{\mathsf{N_R}-1}\psi(\mathbf x,\mathbf y_r)
       \Bigr)
     - \sum_{\mathbf x\in\boldsymbol{\chi}_b^j}\mathsf{N_R} \log \mathbf{L}_{\mathsf{x}}
   \Biggr],$}
\end{align}
with the kernel function :
\begin{equation}
    \psi(\mathbf{x}, \mathbf{y}_r) \coloneqq -\frac{\alpha\left\|\mathbf{x}\right\|^2}{\mathbf L_\mathsf x}+\mathbf{\beta}_x\left|\mathbf{x}^\dag \mathbf{y}_r\right|^2+\frac{2 \sqrt{\alpha}}{\mathbf L_\mathsf x}\left|\mathbf{x}^\dag\mathbf{y}_r\right|\nonumber .
\end{equation}
It's also worth noting that the max-log metric performs nearly as well as the accurate log-based metric since Gray-mapped constellations are in use. The max-log metric appears to have a somewhat minimal impact on receiver performance when operating with low modulation orders.
\begin{remark}
In equation (\ref{eqn:maxlog_llr_simo}), many terms can be omitted when the magnitude of $\mathbf{x}$ remains constant, as is the case for BPSK or QPSK modulations. Additionally, in the presence of strong line-of-sight (LOS) channels, the quadratic terms in equation (\ref{eqn:maxlog_llr_simo}) can also be skipped.
\end{remark}
\begin{corollary}
  Observing the structure of the metric and considering the absence of overlap between the data and the DMRS symbols, we can easily see that the channel estimate is part of the metric.
  By modeling  $\mathbf x$ as a composite signal comprising both data and DMRS in an interleaved fashion within a common OFDM symbol, this yields: $\mathbf x = \mathbf{x^{(\mathsf p)}} + \mathbf{x^{(\mathsf d)}} $
  where  $d$ and $p$ are subscripts representing data, DMRS components, respectively. Hence, we can
  highlight $\hat{\mathbf h}^{\mathrm{LS}}$ through the metric expression:
  \begin{equation}\label{eqn:ce_in_metric}
    \resizebox{0.42\textwidth}{!}{$
      \begin{aligned}
  \left|{\mathbf x}^\dag{\mathbf y_r}\right|
  &=\left|\underbrace{{\mathbf x^{(\mathsf p)}}^\dag\mathbf{y}^{(\mathsf p)}_r}_{\text {channel estimate}}  + \cancel{\underbrace{{\mathbf x^{(\mathsf p)}}^\dag\mathbf{y}^{(\mathsf d)}_r}_{\text {0}}} + \ \cancel{\underbrace{{\mathbf x^{(\mathsf d)}}^\dag\mathbf{y}^{(\mathsf p)}_r}_{\text {0}}} + \ {\mathbf x^{(\mathsf d)}}^\dag\mathbf{y}^{(\mathsf d)}_{r}\right|
  \\&=\left|\underbrace{{\mathbf x^{(\mathsf p)}}^\dag\mathbf{y}^{(\mathsf p)}_r}_{\text {channel estimate}}  + \ {\mathbf x^{(\mathsf d)}}^\dag\mathbf{y}^{(\mathsf d)}_{r}\right|=\left|N_p \hat{\mathrm h}^{{\text{\tiny{LS}}}}_r +  \ {\mathbf x^{(\mathsf d)}}^\dag\mathbf{y}^{(\mathsf d)}_{r}\right| ,
  \end{aligned}
  $}
\end{equation}
\end{corollary}

where $\mathbf {x^{(\mathsf p)}}^\dag \mathbf y^{(\mathsf p)}_{r}=\left(\mathbf{x^{(\mathsf p)}}^\dag\mathbf x^{(\mathsf p)}\right) \hat{\mathrm h}_r^{{\text{\tiny{LS}}}}=\|\mathbf{x^{(\mathsf p)}}\|^2\hat{\mathrm h}_r^{{\text{\tiny{LS}}}}= N_p\rho\hat{\mathrm h}_r^{{\text{\tiny{LS}}}}$ ,
$\rho$ is the reference signal  power and is typically normalized to unity. The channel estimate $\hat{\mathrm{h}}_r^{\text{\tiny{LS}}}$ is obtained via a joint {\em least-squares} (LS) channel estimation using averaging or smoothing over the number of reference signals within the channel coherence (namely here, the coherence bandwidth). 
In general, the channel estimation procedure will work as usual, and the resulting channel estimate is fed into the newly derived soft LLR metric considered in proposition~\ref{prop:4.1}.
\subsubsection{Joint Estimation and Detection}
For the case of polar or LDPC coded data, there is a convincing motivation to divide the coded streams into smaller blocks for detection for some complexity reasons. Assuming an ideal interleaving scenario with perfect CSI, detection can be performed on individual modulated symbols. However, in the presence of the unknown CSI , and considering joint estimation and detection, where interleaved DMRS and data symbols are considered, we need to deal with short blocks that contain both data or modulated symbols and DMRS symbols. Hence, the received and transmitted signal sets are subdivided into smaller block segments, forming detection windows. Phrased directly, the newly proposed metric is applied to each group of symbols, or the so-called detection window. Noting that, in this proposed detection approach, we do not restrict the observation interval of a particular coded bit to the symbol in which it is conveyed.

\begin{proposition}{\em [Block-wise joint estimation - detection.]}\label{prop:4.2}~\\
The proposal consists of applying an advanced joint estimation-detection per block of $M=4$ modulated symbols forming a detection window to generate bit-level LLRs.
Unlike conventional symbol-by-symbol detection in BICM systems where the observation for a particular coded bit is limited to the symbol in which it is conveyed, the proposed block-wise detection approach fundamentally leverages the statistical dependencies between adjacent symbols. In other respect, the LLR for a given coded bit incorporates information from all symbols within the detection window, rather than being constrained to its host symbol alone, thereby improving detection reliability.
\end{proposition}
To illustrate this statement, let $\mathbf{x}^{(\mathsf{d})} = \{\mathrm x^{(\mathsf d)}_n\}_{n=1}^M \in \boldsymbol{\chi}^ M$ be the candidate symbol vector corresponding to a block of $M$ symbols (with $ M=4$ in our case).
Let the received frame be denoted by \(\mathbf{Y} \in \mathbb{C}^{N \times \mathsf{N_R}}\), which contains both data and reference signals. We define the data-carrying positions \((\mathsf{d})\) are divided into \(B\) non-overlapping blocks of size \(M\), such that:
\(
\mathsf{d} = \biguplus_{b=1}^B \mathcal{D}_b, \quad \text{with } |\mathcal{D}_b| = M.
\)
For each block \(b \in \{1,\dots,B\}\), we extract the corresponding \textit{block data observation matrix} from the received signal: \(
\mathbf{Y}_{b}^{(\mathsf{d})} \in \mathbb{C}^{M \times \mathsf{N_R}},
\) which gathers all received data symbols at the \(M\) subcarriers in \(\mathcal{D}_b\) across all \(\mathsf{N_R}\) antennas.
Each block $\mathbf{Y}_{b}^{(\mathsf{d})}$ is processed independently to compute the bit-wise likelihoods over the candidate symbol vectors. More precisely, unlike symbol-wise detection schemes, the proposed block-wise detection computes the LLRs of coded bits by evaluating joint symbol hypotheses over a detection window of size \(M\). This involves assessing all possible candidate vectors \(\mathbf{x}^{(\mathsf{d})} \in \boldsymbol{\chi}^M\), and partitioning them into two subsets ((i.e., grouping vectors where the bit is $0$ vs. $1$) based on the value of the \(j\)-th coded bit. Consequently, each LLR  \(\Lambda^j\left(\cdot \right)\) is derived by marginalizing the block-wise likelihood function.

Considering QPSK modulation and detection windows of $M=4$, each jointly detected symbol block has $|\boldsymbol{\chi}|^M$ = 256 possible candidate symbol vectors in the search space, where each candidate $\mathbf{x}^{(\mathsf{d})} \in \boldsymbol{\chi}^M$ represents a distinct symbol vector. In other terms, the detector exhaustively searches over all  256 possible symbol vector hypotheses to compute the LLRs.

The proposal is conceptually detailed in Algorithm~1.
\begin{figure}[ht]\centering
  \includegraphics[width=1\linewidth]{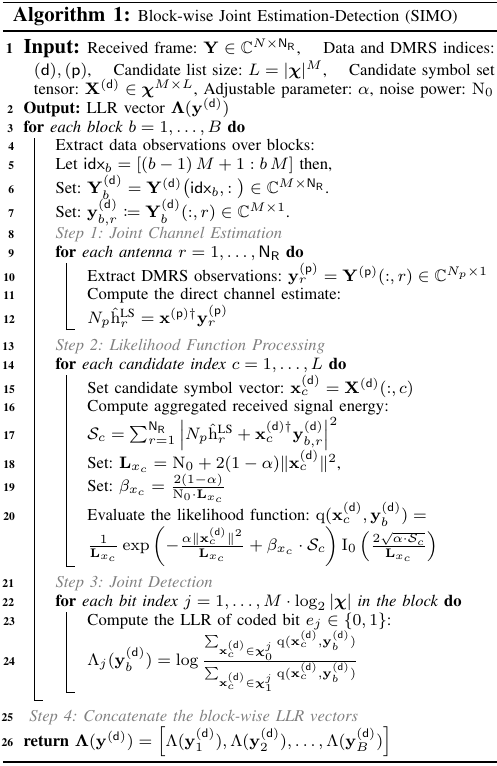}
      \label{algo:JED_SIMO}
      \vspace{-6mm}
\end{figure}

\subsection{\texorpdfstring{$(\mathsf{N_T} \times\mathsf{N_R})$} \ ~MIMO Rayleigh Block Fading Channel}
Consider a SU-MIMO transmission model with $\mathsf{N_T}$ transmit and $\mathsf{N_R}$ receive antennas operating over a quasi-static, flat-fading channel. We assume no {\em inter-symbol interference} (ISI) and consider a time-invariant configuration, making it feasible to use the standard baseband complex-valued representation.
Let $\mathrm{H}_{t,r}$ represent the complex-valued path gain, serving as the fading coefficient from transmit antenna $t$ to receive antenna $r$. At any given time instance, the complex-valued signal $\mathbf x\{\mathrm x_1, \mathrm x_2, \ldots, \mathrm x_{\mathsf{N_T}}\mid  \mathbf x \in \mathbb{C}^{1\times\mathsf{N_T}}\}$ are transmitted through the $\mathsf{N_T}$ respective antennas. The received signal $\mathbf{y}$ belongs to the complex vector space $\mathbb{C}^{1\times\mathsf{N_R}}$. $\mathbf{z} \sim \mathbb{C}^{1\times\mathsf{N_R}}$  is the additive white Gaussian noise with independent real and imaginary components, each having a variance of $\sigma^2$ in every dimension, and the channel matrix is denoted by $\mathbf{H} \in \mathbb{C}^{\mathsf{N_T} \times \mathsf{N_R}}$.
The MIMO channel model is succinctly expressed as
\begin{equation}\label{eqn:sysmodel0}
\mathbf{y}=\mathbf{x}\mathbf{H}+ \mathbf{z}.
\end{equation}
However, the system model given in (\ref{eqn:sysmodel0}), which describes transmission during a single symbol interval, can be extended to multiple consecutive transmissions. Consider \(N\) consecutive transmitted vectors \(\{\mathbf{x}_1, \mathbf{x}_2, \ldots, \mathbf{x}_N\}\), where \(N\) denotes the total number of symbol intervals.\\
For notational convenience, we organize these vectors into matrices as follows:
 $\mathbf{X}=\left[\mathbf{x}_1, \mathbf{x}_2, \ldots, \mathbf{x}_N\right]^\top \in \mathbb{C}^{N \times \mathsf{N_R}}, \quad \mathbf{Y}=\left[\mathbf{y}_1, \mathbf{y}_2, \ldots, \mathbf{y}_N\right]^\top\in \mathbb{C}^{N \times \mathsf{N_R}}, \quad \mathbf{Z}=\left[\mathbf{z}_1, \mathbf{z}_2, \ldots, \mathbf{z}_N\right]^\top \in \mathbb{C}^{N \times \mathsf{N_R}}$.
 The MIMO channel model can then be compactly expressed as
\begin{equation}\label{eqn:sysmodel_matrix}
\mathbf{Y} = \mathbf{X} \mathbf{H} + \mathbf{Z},
\end{equation}
where \(\mathbf{H}\) is assumed to be constant over the \(N\) symbol intervals or the $N$-symbol block and changes independently across blocks. We adopt a wide assumption regarding $\mathbf H \sim \mathbb C\mathcal N(0, \mathbf I)$, which is that its entries, $\mathrm{H}_{t,r}$, are statistically independent for the sake of simplicity.
Accordingly, the  complex-valued fading coefficients $\mathrm{H}_{t,r}$ can be treated as independent zero-mean complex Gaussian random variables with unit variance. Therefore, the MIMO channel model can be referred as the {\em identically and independently distributed} \textit{Rayleigh fading channel}, or more precisely, the Rayleigh block-fading.
Both the fading coefficients and the noise follow complex Gaussian distributions.
%
%
%
Thus, conditioned on the transmitted signal $\mathbf X$, the received signals are jointly complex Gaussian. In other terms, the received signal is zero mean $\mathbb E\{\mathbf Y|\mathbf X\}=0$, circularly symmetric complex Gaussian with a $N\times N$ covariance matrix $\mathbf \Phi_\mathsf Y$, concretely.

Hence, the likelihood function or conditional probability density is simply given by :
\begin{equation}
  p\left(\mathbf Y| \mathbf X\right) \coloneqq \frac{ \exp\left(-\operatorname{tr}\left\{ \mathbf{Y}\mathbf{\Phi}_{\mathsf Y}^{-1} \mathbf{Y}^\dag\right\}\right)}{\pi^{N\times \mathsf{N_R}} \operatorname{det}^{\mathsf{N_R}}\left(\mathbf{\Phi}_{\mathsf Y}\right)}\ .
 \end{equation}
 We will proceed by following the steps below to derive the detection metric.
%
%
Consequently, to determine the formulation of the covariance matrix, $\mathbf{\Phi}_{\mathsf Y}$, we shall invoke the subsequent theorem \cite[Sec. 2, Th. 2]{Gallager2008}, stating that  for any  \(\mathbf{A}\) an arbitrary \(M \times N\) complex matrix, and let \(\mathbf{R} = \mathbf{AW}\), where $\mathbf{W} \sim \mathbb C \mathcal{N}(\mathbf{0}, \mathbf{I}_M)$, meaning  that $W_1, \ldots, W_M$ are independent and identically distributed with independent real and imaginary parts, then:
      \(
      \mathbf{\Phi} = \mathbb{E}[\mathbf{AW} \mathbf{W}^\dag \mathbf{A}^\dag] = \mathbf{AA}^\dag.
      \)  Therefore, \(\mathbf{R} \sim \mathbb C \mathcal{N}(\mathbf{0}, \mathbf{AA}^\dag)\).

Stated directly, the covariance matrix can then be expressed as follows:
 \begin{equation}
     \resizebox{0.43\textwidth}{!}{$
 \begin{aligned}
 \mathbf{\Phi}_{\mathsf Y} &\coloneqq \mathbb E\{\mathbf{Y} \mathbf{Y}^\dag\},\\
 &=\mathbb E\{\left(\mathbf{XH} + \mathbf{Z}\right)\left(\mathbf{XH} + \mathbf{Z}\right)^\dag\}= \mathbb E\{\mathbf{XH}\mathbf{H}^\dag\mathbf{X}^\dag\}+\mathbb E\{ \mathbf{Z}\mathbf{Z}^\dag\},\\
 &=\mathbf{X}~\mathbb E\{\mathbf{H}\mathbf{H}^\dag\}~\mathbf{X}^\dag+\mathbb E\{ \mathbf{Z}\mathbf{Z}^\dag\}
  =\mathbf{X}\mathbf{X}^\dag+ \mathrm {N}_0\mathbf I\ .
 \end{aligned}
 $}
 \end{equation}
This expression of the covariance matrix is commonly encountered in the literature \cite{Marzetta1999} \cite{Hochwald2000}.

Next, the determinant of $\mathbf{\Phi}_{\mathsf Y}$ is shown to be:
  \begin{equation}
 \operatorname{det} \mathbf{\Phi}_{\mathsf Y} =  \operatorname{det}\left(\mathrm {N}_0\mathbf{I} +  \mathbf{X}\mathbf{X}^\dag\right)
=\prod_{n=1}^{N} \left(\mathrm{N}_0 + \lambda_n\right),
 \end{equation}
where \(\lambda_n\) are the eigenvalues of \(\mathbf{X} \mathbf{X}^\dagger\).
Furthermore, the covariance matrix $\mathbf{\Phi}$ involves the addition of two matrices, making it amenable to the application of matrix inversion lemmas such as the Sherman–Morrison–Woodbury formula, or more generally, the Woodbury matrix identity \cite{Woodbury1950}, to compute the inverse of the covariance matrix $\mathbf{\Phi}_{\mathsf{Y}}^{-1} = \left(\mathrm{N}_0 \mathbf{I} + \mathbf{X}\mathbf{X}^\dag\right)^{-1}$.
 \begin{equation}
     \resizebox{0.43\textwidth}{!}{$
 (\mathbf A+\mathbf U \mathbf C \mathbf V)^{-1}\coloneqq\mathbf A^{-1}-\mathbf A^{-1} \mathbf U\left(\mathbf C^{-1}+\mathbf{V A}^{-1} \mathbf U\right)^{-1} \mathbf {V A}^{-1},
 $}
 \end{equation}
 where $\mathbf A$, $\mathbf U$, $\mathbf C$, and $\mathbf V$ are matrices with  comfortable dimensions: $\mathbf A$ is a $n\times n$ matrix, $\mathbf C$ is a $k\times k$ matrix, $\mathbf U$ is a $n\times k$ matrix, and $\mathbf V$ is a $k\times n$ matrix.


 Saying $  \mathbf A= \mathrm {N}_0 \mathbf I, \quad \mathbf C =\mathbf I, \quad \mathbf U= \mathbf X, \quad \mathbf V= \mathbf X^\dag$, then,
\begin{equation}
    \resizebox{0.43\textwidth}{!}{$
\begin{aligned}
\mathbf \Phi^{-1} &\coloneqq \left (\mathbf A+\mathbf U \mathbf C \mathbf V\right)^{-1},\\&
=\mathrm {N}_0^{-1}\mathbf I- \mathrm {N}_0^{-1}\mathbf  X
  \left[\mathrm {N}_0 \mathbf I + \mathbf X^\dag \mathbf X \right]^{-1}\mathbf X^\dag,\\&
  =\mathrm {N}_0^{-1}\mathbf I- \mathrm {N}_0^{-1}\mathbf X \mathbf D\mathbf X^\dag, \quad \text{ where } \mathbf D= \left[\mathrm {N}_0 \mathbf I + \mathbf X^\dag \mathbf X \right]^{-1}.
\end{aligned}
$}
\end{equation}

The likelihood function $\mathrm q\left( \mathbf X, \mathbf Y \right)= p\left(\mathbf Y| \mathbf X\right)$ can be stated as follows:
 \begin{equation}\label{eqn:mimo_lf}
     \resizebox{0.42\textwidth}{!}{$
 \begin{aligned}
   \mathrm q\left( \mathbf X, \mathbf Y \right) &=
  \tfrac{1}{\mathbf L_{\mathsf X}}\exp\left(-\operatorname{tr}\left\{ \mathbf{Y}^\dag\mathbf{\Phi}_{\mathsf Y}^{-1} \mathbf{Y}\right\}\right)\\&
 = \tfrac{1}{\mathbf L_{\mathsf X}}\exp\left(-\operatorname{tr}\left\{ \mathbf{Y}^\dag\left(\frac{1}{\mathrm {N}_0}\mathbf I- \frac{1}{\mathrm {N}_0}\mathbf X \mathbf D\mathbf X^\dag \right) \mathbf{Y}\right\}\right),
  \end{aligned}
  $}
 \end{equation}
 where $\mathbf L_\mathsf{X}= \pi^{N \times \mathsf{N_R}}\operatorname{det}^{\mathsf{N_R}}\left(\mathrm {N}_0\mathbf I +  \mathbf{X}\mathbf{X}^\dag\right)$.\\ Ignoring the multiplicative terms independent of $\mathbf X$, (\ref{eqn:mimo_lf}) reduces to:
 \begin{equation}
 \begin{aligned}
   \mathrm q\left(\mathbf X, \mathbf Y\right) \propto
 \frac{1}{\mathbf L_{\mathsf X}} \exp\left(\frac{1}{\mathrm {N}_0} \operatorname{tr}\left\{\left(\mathbf X^\dag\mathbf{Y}\right)^\dag \mathbf D\left(\mathbf X^\dag \mathbf{Y}\right)\right\}\right).
  \end{aligned}
 \end{equation}
 As described in (\ref{eqn:ce_in_metric}), we can incorporate the channel estimate into the metric to take the full merit of the JED principle.
 For this purpose, we simply rewrite $\mathbf X =  \mathbf X^{(\mathsf d)} + \mathbf X^{(\mathsf p)}$.
 Then, we can reveal $\widehat{\mathbf H}_{\mathrm{LS}}$ in the metrics:
 \begin{equation}\label{joint_estim_mimo}
\resizebox{0.42\textwidth}{!}{$
 \begin{aligned}
 \mathbf X^\dag \mathbf Y = \underbrace{{\mathbf X^{(\mathsf p)}}^\dag\mathbf Y^{(\mathsf p)} }_{\text{channel estimate  }}  + \ {\mathbf X^{(\mathsf d)}}^\dag\mathbf Y^{(\mathsf d)}= \mathbf C_{\mathsf p}\widehat{\mathbf H}_{\mathrm{LS}} + {\mathbf X^{(\mathsf d)}}^\dag\mathbf Y^{(\mathsf d)}\ ,
 \end{aligned}
 $}
 \end{equation}
 where $\mathbf C_{\mathsf p} ={\mathbf {X}^{(\mathsf p)}}^\dag\mathbf X^{(\mathsf p)}$ given that
 $\widehat{\mathbf H}_{\mathrm{LS}}  = \frac{{\mathbf {X}^{(\mathsf p)}}^\dag\mathbf Y^{(\mathsf p)}}{{\mathbf {X}^{(\mathsf p)}}^\dag\mathbf X^{(\mathsf p)}}$. This channel estimate is obtained via a joint  least-squares (LS) channel estimation using averaging or smoothing over the number of dimensions exhibiting channel coherence.
\begin{proposition}\label{prop:4.3} {\em [A novel soft-likelihood metric for MIMO.]}~\\
 Consistent with the above derivation steps, the proposed likelihood function for joint estimation - detection over MIMO Rayleigh block fading channel can be subsequently formulated as follows :
\begin{equation}
  \resizebox{0.42\textwidth}{!}{$
 \begin{aligned}\label{eqn:_jed_lf_mimo}
   \mathrm q(\mathbf{X}, \mathbf{Y}) & =\frac{1}{\mathbf L_\mathsf X} \exp \left(\frac { 1 } { \mathrm N _ { 0 } } \operatorname { t r } \left\{\left(\mathbf C_{\mathsf p} \widehat{\mathbf{H}}_{\mathrm{LS}}+{\mathbf{X}^{(\mathsf d)}}^\dag \mathbf{Y}^{(\mathsf d)}\right)^\dag \right.  \right. \\
   &\hspace{6.5em}\left. \left.\mathbf{D}\left(\mathbf C_{\mathsf p}\widehat{\mathbf{H}}_{\mathrm{LS}}+{\mathbf{X}^{(\mathsf d)}}^\dag \mathbf{Y}^{(\mathsf d)}\right)\right\} \right).
  \end{aligned}
  $}
 \end{equation}
\end{proposition}
  Then, the likelihood of the coded bit $e_j$ s.t. $b \in \{0,\ 1\}$ is given  by
 \begin{equation}\label{eqn:llr_mimo_rbf}
 \mathrm q\left(e_j(\mathbf{X})=b ,\mathbf Y\right) = \displaystyle \sum_{\mathbf X \ \in \ \boldsymbol{\chi}_b^j} \mathrm q\left(\mathbf X, \mathbf Y\right).
 \end{equation}
 The LLR bit metric  for the $j-th$  bit in BICM receiver  is
 \begin{equation}
 \begin{aligned}\label{eqn:llr_jed_mimo}
 \Lambda^j\left(\mathbf Y\right)&=\log \frac{\mathrm q\left(e_j(\mathbf X)=0,\mathbf{Y}\right)}{\mathrm q\left(e_j(\mathbf X)=1,\mathbf{Y}\right)}.
 \end{aligned}
 \end{equation}
To ease the process of implementing such a LLR bit metric in (\ref{eqn:llr_jed_mimo}), one may use its {\em max-log approximation} version given in (\ref{eqn:maxlog_llr_mimo}).
\vspace{-1em}
\begin{equation}\label{eqn:maxlog_llr_mimo}
    \resizebox{0.42\textwidth}{!}{$
 \begin{aligned}
\Lambda^j(\mathbf Y)
\;=\;
\sum_{b\in\{0,1\}}(-1)^{b}\,
\Biggl[
  \max_{\mathbf X\in\boldsymbol{\chi}_b^j}
    \Gamma\bigl(\mathbf X,\mathbf Y\bigr)
  \;-\;
  \sum_{\mathbf X\in\boldsymbol{\chi}_b^j}\!\log \mathbf L_\mathsf X
\Biggr].
 \end{aligned}
 $}
\end{equation}
with the score function or max-log decision metric
  \[\resizebox{0.50\textwidth}{!}{$
\Gamma\bigl(\mathbf X,\mathbf Y\bigr)
\;\coloneqq\;
\frac{1}{\mathrm {N}_0}\;
\operatorname{tr}\!\Bigl\{\,
\bigl(\mathbf C_{p}\,\widehat{\mathbf H}_{\mathrm{LS}}
       +{\mathbf X}^{(d)\,\dagger}\,\mathbf Y^{(d)}\bigr)^{\!\dagger}
\;\mathbf D\;
\bigl(\mathbf C_{p}\,\widehat{\mathbf H}_{\mathrm{LS}}
       +{\mathbf X}^{(d)\,\dagger}\,\mathbf Y^{(d)}\bigr)
\Bigr\}\,.
$}\]
\begin{remark}
For a MIMO system with $\mathsf{N_T}$ transmit antennas employing block-wise detection (block size $M$) and QPSK modulation ($|\boldsymbol{\chi}|=4$), each antenna's symbol vector is selected from $|\boldsymbol{\chi}|^M$ possible candidates. The joint search space across all $\mathsf{N_T}$ antennas then expands to $\big(|\boldsymbol{\chi}|^M\big)^\mathsf{N_T}$ possible combinations - computationally expensive but optimal (ML joint detection). However, under the assumption of no spatial correlation, the search space can be processed independently for each of the $\mathsf{N_T}$ streams. In this case, each transmit stream  is detected independently without inter-antenna joint processing. This stream-wise implementation reduces the computational complexity from exponential to linear scaling with respectt to $\mathsf{N_T}$, requiring only $\mathsf{N_T} \times |\boldsymbol{\chi}|^M$ total candidates. This per-stream block-wise joint detection approach demonstrates somewhat significantly reduced complexity compared to full joint detection (i.e., joint-stream block-wise detection).
 \end{remark}

Algorithm~2 outlines this stream-wise implementation of block-wise JED for spatially multiplexed MIMO systems.
\vspace{-6mm}
\begin{figure}[ht]\centering
  \includegraphics[width=1\linewidth]{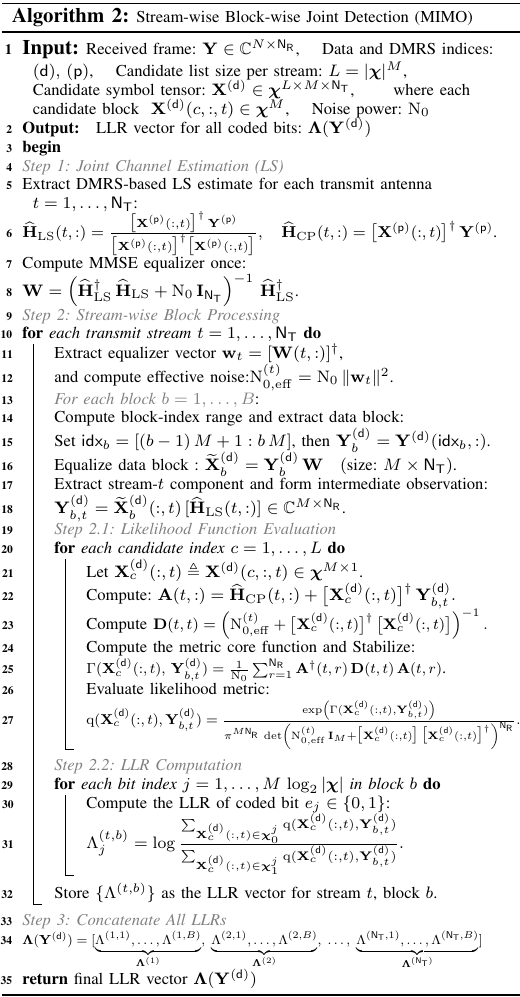}
      \label{algo:JED_MIMO}
      \vspace{-6mm}
\end{figure}

\section{Numerical Results}
For illustrative purposes, we consider two distinct configurations: $(1\times 4)$ SIMO BICM,  and  $(4 \times 4)$  MIMO BICM. The subsequent figures illustrate the performance of BIPCM/BILCM with joint estimation and detection using the proposed block-wise joint estimation–detection metrics under various channel conditions. The evaluation highlights performance differences across three scenarios: \emph{Perfect CSI} or ideal or  coherent receiver, \emph{No CSI} JED using the block-wise($M=4$) joint detection approach, and \emph{No CSI} Conv, using conventional symbol-by-symbol detection.

The simulations employ NR POLAR and NR LDPC coding schemes paired with QPSK modulation. The transmission involves a transport block of 48 bits. The resource allocation procedure uses a single OFDM symbol with 48 resource elements spread over 4 PRBs (comprising 32 REs for data component and fewer REs for DMRS component).
\subsection{Performance Analysis}
\subsubsection{SIMO Non Coherent Fading channel}
\begin{figure*} [!ht]
    \centering
  \subfloat[\scriptsize {SIMO LOS channel, 5G polar BICM. }\label{fig:simo_polar_los}]{
    \resizebox{0.40\linewidth}{!}{\pgfplotsset{
    tick label style={font=\footnotesize},
    label style={font=\footnotesize},
    legend style={nodes={scale=0.6, transform shape}, font=\footnotesize, legend cell align=left},
    every axis/.append style={line width=0.5pt}
}


\begin{tikzpicture}[spy using outlines=
      	{circle, magnification=3, connect spies}]
              \begin{semilogyaxis}[xlabel=\footnotesize{SNR (dB)},
            ,ylabel=\footnotesize{BLER},
            legend pos=north east,
           legend columns=2,
            legend style={at={(0.5,-.22)},anchor=north},
            grid=both,
            ytick={1e-1,1e-2,1e-3,1e-4},
             minor tick num=3,
            width=6.8cm,
            height=5cm,
            ymax=0.7, ymin=0.00005,xmin=-4, xmax=8
    ]

  \addplot+[smooth,teal, line  width=0.8pt, mark options={scale=1, fill=teal, solid},mark=otimes]
    table[row sep=crcr]{%
  -2.5	0.130952380952381\\
   -1.5	0.0188172043010753\\
  -0.9	0.00252474247626742\\
  0	0.00012\\
  };

%

\addplot+[smooth,densely dotted, black, line width=0.8pt, mark options={scale=0.7, fill=black, solid},mark=square*]
table[row sep=crcr]{%
-0.5	0.0831255195344971\\
1	0.0079\\
2.5	0.0002\\
};

\addplot+[smooth,densely dotted, gray, line width=0.8pt, mark options={scale=0.7, fill=gray, solid},mark=otimes]
table[row sep=crcr]{%
2	0.103305785123967\\
3.5	0.0161864681126578\\
5	0.0026\\
6.5	0.0004\\
};

\addplot+[smooth, dashed, violet,line  width=0.8pt, mark options={scale=1, fill=violet, solid},mark=otimes]
table[row sep=crcr]{%
-1	0.088809946714032\\
0.5	0.006\\
1	0.00205372545798078\\
1.4	0.0008\\
1.8	0.0002\\
};


\addplot+[smooth,Goldenrod, line width=0.8pt, mark options={scale=0.7, fill=Goldenrod, solid},mark=asterisk]
table[row sep=crcr]{%
-1.5	0.114678899082569\\
-0.5	0.0137400384721077\\
1	0.0005\\
};
%

\addplot+[smooth, solid, green,  line width=0.8pt, mark options={scale=1, fill=green, solid},mark=*]
table[row sep=crcr]{%
-2	0.107991360691145\\
-1.5	0.0484261501210654\\
-0.5	0.0037\\
0.5	0.0002\\
};

\addplot+[smooth, solid, red,line  width=0.8pt, mark options={scale=1, fill=red, solid},mark=]
table[row sep=crcr]{%
-2.5	0.0584257447239825\\
-2	0.011142474157713\\
-1.5	0.000588429070017758\\
-1	2.98435346820854e-06\\
-0.5	0\\
};


\addplot+[smooth, densely dotted, cyan, line  width=0.8pt, line  width=0.8pt, mark options={ scale=1, fill=cyan, solid},mark=]
table[row sep=crcr]{%
1.5	0.0661898878566984\\
2	0.00755567198829342\\
2.5	0.000380799694928626\\
3	1.42626375637674e-05\\
};

\coordinate (spypoint) at  (axis cs:0.6,0.01);    
\coordinate (magnifyglass) at (axis cs:1.7,0.4); 

\coordinate (spypoint2) at  (axis cs:-1.5,0.09);    
\coordinate (magnifyglass2) at (axis cs:-1, 0.2); 

\coordinate (spypoint3) at  (axis cs:4,0.01);    
\coordinate (magnifyglass3) at (axis cs:5, 0.1); 

\end{semilogyaxis}


\node[
    fill=white,
    fill opacity=0.5,
    text=black,
    font=\tiny,
    rounded corners=2pt,
    align=center,
    anchor=west
]
at (magnifyglass) {JED vs Conv \\ LS CE+Avg};

\draw[-, thick, gray] (magnifyglass) -- (spypoint);


\draw[
    draw=gray,
    line width=0.5pt,
    rotate around={45:($(spypoint)$)}
]
($(spypoint)$) ellipse [x radius=0.29, y radius=0.15];

\node[
    fill=white,
    fill opacity=0.5,
    text=black,
    font=\tiny,
    rounded corners=2pt,
    align=center,
    anchor=south
]
at (magnifyglass2) {DMRS Power \\ Boosting};
%
\draw[-, thick, gray] (magnifyglass2) -- (spypoint2);

\draw[
    draw=gray,
    line width=0.5pt,
    rotate around={45:($(spypoint2)$)}
]
($(spypoint2)$) ellipse [x radius=0.20, y radius=0.15];

\node[
    fill=white,
    fill opacity=0.5,
    text=black,
    font=\tiny,
    rounded corners=2pt,
    align=center,
    anchor=west
]
at (magnifyglass3) {Conv.\\ LS CE+Interpol.};

\draw[-, thick, gray] (magnifyglass3) -- (spypoint3);

\draw[
    draw=gray,
    line width=0.5pt,
    rotate around={45:($(spypoint3)$)}
]
($(spypoint3)$) ellipse [x radius=0.10, y radius=0.10];

\end{tikzpicture}}
  }
  \subfloat[\scriptsize {SIMO LOS channel, 5G LDPC BICM.}\label{fig:simo_ldpc_los}]{%
     \resizebox{0.40\linewidth}{!}{\pgfplotsset{
    tick label style={font=\footnotesize},
    label style={font=\footnotesize},
    legend style={nodes={scale=0.6, transform shape}, font=\footnotesize, legend cell align=left},
    every axis/.append style={line width=0.5pt}
}


\begin{tikzpicture}[spy using outlines=
      	{rectangle, magnification=4, connect spies}]
              \begin{semilogyaxis}[xlabel=\footnotesize{SNR (dB)},
            ,ylabel=\footnotesize{BLER},
            legend pos=north east,
           legend columns=2,
            legend style={at={(0.5,-.22)},anchor=north},
            grid=both,
             minor tick num=9,
            width=6.8cm,
            height=5cm,
              ytick={1e-1,1e-2,1e-3,1e-4,1e-5,1e-6},
            ymax=0.7, ymin=0.0000006,xmin=-4, xmax=12
    ]

   \addplot+[smooth,teal, line  width=0.8pt, mark options={scale=1, fill=teal, solid},mark=otimes]
    table[row sep=crcr]{%
    0.5	0.21978021978022\\
    1.8	0.00108089411561243\\
    2.8	1.5e-05\\
    3.4	2e-06\\
  };

  \addplot+[smooth,densely dotted, black, line width=0.8pt, mark options={scale=0.7, fill=black, solid},mark=square*]
  table[row sep=crcr]{%
  1.6	0.146842878120411\\
  2.8	0.00482067103740841\\
  4	0.000117414302236155\\
  5.2	2e-06\\
  };

\addplot+[smooth, densely dotted, gray, line  width=0.8pt, mark options={ scale=1, fill=gray, solid},mark=otimes]   table[row sep=crcr]{%
 4	0.0742390497401633\\
 9.5	3.6e-05\\
11.5	1e-06\\
};

\addplot+[smooth, dashed, violet,line  width=0.8pt, mark options={scale=1, fill=violet, solid},mark=otimes]
table[row sep=crcr]{%
1	0.115740740740741\\
2	0.00597907324364723\\
3	0.000235698965281542\\
4	9e-06\\
};

\addplot+[smooth,Goldenrod, line width=0.8pt, mark options={scale=0.7, fill=Goldenrod, solid},mark=asterisk]
table[row sep=crcr]{%
1	0.0863557858376511\\
2	0.00236893847866771\\
3	8.4e-05\\
4	1e-06\\
};

\addplot+[smooth, solid, green,  line width=0.8pt, mark options={scale=1, fill=green, solid},mark=*]
table[row sep=crcr]{%
1	0.0635324015247776\\
2	0.00135442626503413\\
3	5.8e-05\\
4	1e-06\\
};

\addplot+[smooth, solid, red,line  width=0.8pt, mark options={scale=1, fill=red, solid},mark=]
table[row sep=crcr]{%
-2.5	0.0584257447239825\\
-2	0.011142474157713\\
-1.5	0.000588429070017758\\
-1	2.98435346820854e-06\\
};

\addplot+[smooth, densely dotted, cyan, line  width=0.8pt, line  width=0.8pt, mark options={ scale=1, fill=cyan, solid},mark=]
table[row sep=crcr]{%
1	0.267468963044691\\
1.5	0.0661898878566984\\
2	0.00755567198829342\\
2.5	0.000380799694928626\\
3	1.42626375637674e-05\\
3.5	6.59382728205706e-07\\
};

 \coordinate (spypoint_zoom) at (axis cs:1.7,0.01);   
 \coordinate (magnifyglass_zoom) at (axis cs:0,0.00005); 

\coordinate (spypoint) at  (axis cs:3.2,0.001);    
\coordinate (magnifyglass) at (axis cs:3.5,0.1); 

\coordinate (spypoint2) at  (axis cs:-1.5,0.09);    
\coordinate (magnifyglass2) at (axis cs:-1, 0.2); 

\coordinate (spypoint3) at  (axis cs:5.7,0.01);    
\coordinate (magnifyglass3) at (axis cs:7, 0.1); 

\end{semilogyaxis}
        \spy [black, size=0.8cm] on (spypoint_zoom) in node[fill=white, size=1.cm] at (magnifyglass_zoom);


\node[
    fill=white,
    fill opacity=0.5,
    text=black,
    font=\tiny,
    rounded corners=2pt,
    align=center,
    anchor=south
]
at (magnifyglass) {JED vs Conv \\ LS CE+Avg};

\draw[-, thick, gray] (magnifyglass) -- (spypoint);


\draw[
    draw=gray,
    line width=0.5pt,
    rotate around={45:($(spypoint)$)}
]
($(spypoint)$) ellipse [x radius=0.28, y radius=0.15];

%
%

\node[
    fill=white,
    fill opacity=0.5,
    text=black,
    font=\tiny,
    rounded corners=2pt,
    align=center,
    anchor=west
]
at (magnifyglass3) {Conv.\\ LS CE+Interpol.};

\draw[-, thick, gray] (magnifyglass3) -- (spypoint3);

\draw[
    draw=gray,
    line width=0.5pt,
    rotate around={45:($(spypoint3)$)}
]
($(spypoint3)$) ellipse [x radius=0.15, y radius=0.10];

\end{tikzpicture}}
   }
    \vspace{-2mm}
  \\
       \subfloat[\scriptsize {SIMO Rayleigh fading channel, 5G polar BICM.}\label{fig:simo_polar_ray}]{
         \resizebox{0.40\linewidth}{!}{\pgfplotsset{
    tick label style={font=\footnotesize},
    label style={font=\footnotesize},
    legend style={nodes={scale=0.6, transform shape}, font=\footnotesize, legend cell align=left},
    every axis/.append style={line width=0.5pt}
}


\begin{tikzpicture}[spy using outlines=
      	{circle, magnification=3, connect spies}]
              \begin{semilogyaxis}[xlabel=\footnotesize{SNR (dB)},
            ,ylabel=\footnotesize{BLER},
            legend pos=north east,
           legend columns=2,
            legend style={at={(0.5,-.22)},anchor=north},
            grid=both,
             minor tick num=3,
            width=6.8cm,
            height=5cm,
            ymax=0.7, ymin=0.0005,xmin=-1.5, xmax=12
    ]

  \addplot+[smooth,teal, line  width=0.8pt, mark options={scale=1, fill=teal, solid},mark=otimes]
    table[row sep=crcr]{%
  0	0.101010101010101\\
  5	0.00261780104712042\\
  8	0.0002\\
  };

\addplot+[smooth, densely dotted, gray, line  width=0.8pt, mark options={ scale=1, fill=gray, solid},mark=otimes]   table[row sep=crcr]{%
 4	0.116686114352392\\
8	0.0091\\
10	0.0017\\
12	0.0002\\
};

%

\addplot+[smooth, dashed, violet,line  width=0.8pt, mark options={scale=1, fill=violet, solid},mark=otimes]
table[row sep=crcr]{%
2	0.0847457627118644\\
5.5	0.008\\
8.5	0.0008\\
};

\addplot+[smooth,Goldenrod, line width=0.8pt, mark options={scale=0.7, fill=Goldenrod, solid},mark=asterisk]
table[row sep=crcr]{%
1.5	0.0889679715302491\\
3	0.0343288705801579\\
5	0.0083\\
8	0.0008\\
};

%

\addplot+[smooth, solid, green,  line width=0.8pt, mark options={scale=1, fill=green, solid},mark=*]
table[row sep=crcr]{%
1.5	0.0694927032661571\\
4.5	0.01001001001001\\
7.5	0.001\\
};

\addplot+[smooth, dashed, red,line  width=0.8pt, mark options={scale=1, fill=red, solid},mark=]
table[row sep=crcr]{%
 -1	0.0370907843219914\\
-0.5	0.0233971427820133\\
0	0.014560533156261\\
0.5	0.00881198959720277\\
1	0.00549482598471063\\
1.5	0.00319772423413608\\
2	0.00195752949148595\\
2.5	0.00116746048383612\\
3	0.000638313134312618\\
3.5	0.000401458321201706\\
4	0.000273825851137497\\
4.5	0.000167999165263977\\
5	0.000104785275465778\\
};


%
%
%
%

\end{semilogyaxis}

\end{tikzpicture}}
         }
   \subfloat[\scriptsize {SIMO TDL-C channel, 5G Polar BICM}\label{fig:simo_polar_tdl}]{%
      \resizebox{0.40\linewidth}{!}{\pgfplotsset{
    tick label style={font=\footnotesize},
    label style={font=\footnotesize},
    legend style={nodes={scale=0.6, transform shape}, font=\footnotesize, legend cell align=left},
    every axis/.append style={line width=0.5pt}
}


\begin{tikzpicture}[spy using outlines=
      	{circle, magnification=3, connect spies}]
              \begin{semilogyaxis}[xlabel=\footnotesize{SNR (dB)},
            ,ylabel=\footnotesize{BLER},
            legend pos=north east,
           legend columns=2,
            legend style={at={(0.5,-.22)},anchor=north},
            grid=both,
             minor tick num=3,
            width=6.8cm,
            height=5cm,
            ymax=0.7, ymin=0.001,xmin=-2, xmax=11
    ]

  \addplot+[smooth,teal, line  width=0.8pt, mark options={scale=1, fill=teal, solid},mark=otimes]
    table[row sep=crcr]{%
    -1	0.182815356489945\\
    2	0.0285388127853881\\
    5	0.0018\\
  };

\addplot+[smooth, densely dotted, gray, line  width=0.8pt, mark options={ scale=1, fill=gray, solid},mark=otimes]   table[row sep=crcr]{%
2	0.294985250737463\\
5	0.076\\
10	0.002\\
};

\addplot+[smooth, dashed, violet,line  width=0.8pt, mark options={scale=1, fill=violet, solid},mark=otimes]
table[row sep=crcr]{%
1	0.238663484486873\\
3	0.082\\
9	0.002\\
};

\addplot+[smooth,Goldenrod, line width=0.8pt, mark options={scale=0.7, fill=Goldenrod, solid},mark=asterisk]
table[row sep=crcr]{%
0	0.294985250737463\\
3	0.0629722921914358\\
8	0.00345661942620118\\
};
%

\addplot+[smooth, solid, green,  line width=0.8pt, mark options={scale=1, fill=green, solid},mark=*]
table[row sep=crcr]{%
0	0.268096514745308\\
3	0.0504286434694907\\
7	0.0049\\
};

%
%

\end{semilogyaxis}

\end{tikzpicture}}
    }\\
    \vspace{2mm}
     \hspace{4em }\includegraphics[width=0.7\linewidth]{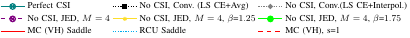}
     \vspace{-1em}
   \caption{Block Error Rate, 48 bits(TBs+CRC), NR POLAR BICM (CRC-aided successive-cancellation list decoder, List length=8), NR LDPC BICM (belief propagation decoder, iteration=30) QPSK modulation, 1 OFDM symbol, 4 PRBs, 48 REs (32 data, 4 DMRS), ($1\times4$) SIMO, vs outer (MC) and inner (RCUs) bounds,  DMRS power Boosting via a scaling factor $\beta$.}
   \label{fig:simo_nocoh_fading_metric}
   \vspace{-6mm}
 \end{figure*}

The results illustrated in Figure~\ref{fig:simo_polar_los} Figure~\ref{fig:simo_polar_ray} delineate the performance of the ($1\times 4$) SIMO BICM systems, with joint estimation and detection, over LOS channel ($\alpha=1$) and rayleigh fading channel ($\alpha=0$), respectively. This evaluation is aiming to discern the performance disparity between the \emph{Perfect CSI} and \emph{No CSI} conditions, particularly in coverage scenarios characterized by low signal-to-noise ratios. Moreover, we explore the most challenging situation where the DMRS density is very low, in order to show how robust the proposed receiver metrics can be compared to the conventional ones. To do this, we consider a single DMRS symbol across the entire coherence bandwidth or, at the very least, across a PRB. In this way, two structures of the conventional receiver are considered: (\emph{No CSI Conv., LS CE + Interpol.}), where least squares channel estimation is followed by linear interpolation, and (\emph{No CSI Conv., LS CE + Avg.}), where channel estimation is performed by averaging over the DMRS symbols for a fairer comparison with the proposed JED receiver.

Figure~\ref{fig:simo_polar_los} illustrates the performance over a pure LOS channel (\(\alpha=1\)) under 5G polar coded modulation. The JED-based receiver achieves a performance gain of 3.75 dB and 0.75 dB compared to the conventional receivers at a BLER threshold of 1\%. However, a gap of approximately 0.5 dB remains between the JED receiver (with \(M=4\)) and the perfect CSI receiver.\\
Similarly, Figure~\ref{fig:simo_ldpc_los} shows the performance under 5G LDPC coded modulation. The JED-based receiver provides a performance gain of 3.5 dB over the conventional receiver (\emph{No CSI Conv., LS CE + Interpolation}) at the same BLER threshold.\\ 
To further highlight the advantage of the JED receiver over the conventional receiver, Figure~\ref{fig:simo_ldpc_los} depicts performance over a pure Rayleigh fading channel (\(\alpha=0\)) under 5G polar coded modulation. Here, the JED-based receiver (\(M=4\)) achieves a gain of 2.75 dB with respect to the conventional receiver (\emph{No CSI Conv., LS CE + Interpolation}). 

 In addition, the Figures also includes finite blocklength bounds, integrating both converse and achievability results for a comprehensive comparative analysis. References are made to the scholarly works \cite{Polyanskiy2010, Martinez2011} detailing the \emph{metaconverse} (MC) and \emph{Random Coding Union} (RCU) bounds employed. Indeed, The RCU provides an achievability bound and gives a practical estimate of how well a communication system can perform at finite blocklength. It estimates the error probability of a random code at a given blocklength, rate, and SNR. Conversely, the meta-converse bound establishes a lower bound on the minimum achievable error probability for any coding scheme. In other words, it complements the finite blocklength achievability bound by setting a benchmark for the best possible performance, ensuring that practical systems do not overestimate their capabilities, which justifies its inclusion in the graphs.\\As shown in Figure~\ref{fig:simo_polar_los}, the Gallager parameter $s$ is numerically optimized (i.e., $s \to s^*$) to maximize the RCU bound and minimize the meta-converse Verdú-Han bound (i.e., the tightest bound), where $s^*$ corresponds to the saddle point. Conversely, in Figure~\ref{fig:simo_polar_ray}, no optimization of the Gallager parameter $s$ is performed; instead, a fixed value of $s = 1$ is chosen, resulting in a more relaxed constraint.

Next, Figure~\ref{fig:simo_polar_tdl} compares the performances under a TDL-C Non-Line-of-Sight (Long Delay Spread = 300 ns, urban macro, Sampling rate $f_s$ = 30.72 MHz) channel configuration with 4 receive antennas. The antenna ports were subjected to independent and identically distributed realizations, with no correlation modeling applied. Indeed, the TDL-C, is a 3GPP reference channel model characterized by its long delay-spread and emphasis for non-MIMO evaluation scenarios \cite{3GPP38901}. It models a multipath-rich environment where no dominant direct path (LOS) is explicitly present. Specifically, the TDL channel, being purely NLOS, behaves like a multi-tap Rayleigh channel where all taps are statistically distributed as independent and random Rayleigh fading. Consequently, we implement the proposed metric by setting $\alpha=0$  in our simulations.
Since the TDL-C channel is frequency selective, channel estimation using the least squares method followed by interpolation is typically required to better track channel fluctuations. However, it is important to recall that for the JED receiver under consideration, no interpolation is performed. Instead, a least-squares (LS) estimation is conducted, followed by averaging over the dimension of DMRS symbols, before incorporating the resulting channel estimate into the soft detection metric.

\begin{figure} [!ht]
     \centering
     \includegraphics[width=1\linewidth]{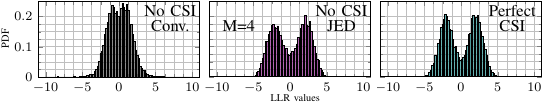}
     \caption{LLR distribution characteristics via histograms at SNR = 0 dB.}
   \vspace{-2mm}
   \label{fig:histo}
 \end{figure}

Furthermore, Figure~\ref{fig:histo} displays LLR distribution characteristics shown via histograms at SNR = 0 dB for the \emph{No CSI} Conv., the proposed \emph{No CSI} JED  ($M=4$), and the \emph{Perfect CSI} receivers. It should be noted that LLR values represent the confidence level in binary decisions (bit = 0 or 1) after demodulation. An LLR value $\approx$ 0 indicates significant uncertainty or ambiguity in the binary decision, whereas values $\gg$ 0 or $\ll$ 0 correspond to high-confidence decisions (bit = 1 or 0, respectively).
The histogram shapes thus reflect the quality of information provided by each receiver. The \emph{No CSI} JED ($M=4$) receiver exhibits a bimodal histogram with peaks near $-3$ and $+3$, showing well-separated LLR values that correspond to more confident and accurate decisions. This strongly suggests effective receiver operation without explicit CSI, further supported by the close similarity between its histogram and that of the ideal receiver.
The discrepancy between the ideal receiver and the proposed one can be explained by the fact that the simulations are conducted under low DMRS density scenarios (i.e., 1 DMRS per PRB). In other respect, only fewer resource elements are allocated to DMRSs. This helps to some extent reduce the additional transmission overhead inherent in dense DMRS-assisted schemes.

Lastly, To fully exploit the performance potential of the JED receiver, we explore extra enhancement strategies. There are two possibilities: (a) DMRS densification, and (b) DMRS power boosting. If a certain sweet pot in terms of DMRS density per PRB cannot be found, increasing DMRS density inherently introduces additional transmission overhead. Therefore, the ideal approach is to prioritize transmission with low DMRS density just to bootstrap the JED receiver, followed by DMRS power boosting. Indeed, DMRS power boosting was extensively discussed in our prior correspondence \cite{sy2023_2}, particularly in scenarios where reference and data symbols are jointly conveyed in common OFDM symbols. Conceptually, envision the signal as comprising a data component and a data-independent component, or pilots, in a frequency-interleaved fashion. To enhance the power of pilot signals within an interleaved set, scaling the power of DMRSs while keeping the data signals unchanged or constant is crucial. Put simply, the boosted transmitted signal, denoted as $\mathbf{x}_{\texttt{boosted}}$, is then defined as $\mathbf{x}_{\texttt{boosted}} = \mathbf{x}^{(\mathsf d)} + \beta \ \mathbf{x}^{(\mathsf p)}$. The adaptive power adjustment procedure is contingent on $\beta$ values and aims to increase the power or strength of the pilot signals within the composite signal. Care should be taken to select an appropriate value for $\beta$ to achieve the desired power augmentation without introducing distortion or signal saturation. To comply with potential radio frequency constraints, $\beta$ must be perfectly calibrated. Consequently, optimal performance enhancement is achieved when $\beta$ is set to $1.75$ (corresponding to a $75\%$ increase in DMRS power w.r.t. its initial value). For instance, Figure~\ref{fig:simo_polar_los} shows that the proposed JED-based receiver delivers an additional gain of approximately $1$ dB, approaching the performance bound of the ideal receiver. Overall, the implications of slightly adjusting the DMRS power within the 3GPP standard are significant. Specifically, it is feasible to allow the {\em user equipment} (UE) to adjust the power allocation between the DMRS and data transmission. This flexibility in adaptive DMRS power adjustment is somewhat transparent to the receiver.

\subsubsection{MIMO Rayleigh Block Fading Channel}
In previous simulations, we considered a low-density DMRS scenario with 1 DMRS per PRB. However, for MIMO cases, we increased the DMRS density to 4 per PRB.
Indeed, channel estimation quality depends on the total number of DMRS resources per antenna stream not just on having orthogonal sequences.
For a ($4\times 4$) MIMO configuration, 1 DMRS per antenna (stream) is insufficient to properly estimate all four channels. Multiple DMRS (e.g., 4) per antenna are required to ensure stable channel estimation. While a single orthogonal DMRS sequence can uniquely identify each antenna’s channel, it does not provide reliable estimation, as it remains sensitive to channel variations, noise, and interference

Figure~\ref{fig:mimo_polar_ray} illustrates the performance over a ($4\times 4$) MIMO Rayleigh block fading channel under 5G polar-coded modulation. The JED-based receiver achieves a performance gain of approximately 2.75 dB compared to the conventional receiver at a BLER threshold of 0.1\%. However, there remains a performance gap of about 1.75 dB between the JED receiver ($M = 4$) and the ideal or \emph{Perfect CSI} receiver

Similarly, Figure~\ref{fig:mimo_LDPC_ray} shows the performance under 5G LDPC-coded modulation. The JED-based receiver provides a gain of approximately 2 dB over the conventional receiver, and  the gap between the JED receiver ($M = 4$) and the perfect CSI receiver is around 1.75 dB.
   %
   %

Therefore, it can be remarkably asserted that the advanced JED-based receiver outperforms the conventional counterpart and demonstrates greater resilience under imperfect channel estimation.
\subsection{Complexity Analysis}
\begin{figure} [!ht]
\centering
   \resizebox{0.75\linewidth}{!}{\includegraphics[width=0.7\linewidth]{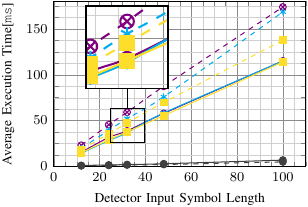}}\\
 \vspace{1mm}
  \hspace{0em }\includegraphics[width=1\linewidth]{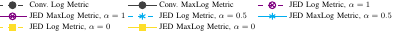}
  \vspace{-2em}
    \caption{Average Time Complexity of the proposed SIMO JED metrics vs Conventional metrics.}
  \label{fig:complexity_simo}
\end{figure}

The complexity of the detection metrics is analysed using {\em Monte Carlo} simulation. The execution time highlights the time elapsed between the input and output of the demodulator, concisely, until the LLRs are generated. It is relevant  to pinpoint the block size range wherein complexity  is relatively low compared to conventional metrics in order to establish a better trade-off between performance and complexity.
Analytically, within a very short block regime(.i.e, small input symbol size), the conventional  and JED receiver metrics will almost close. Considering the JED Metrics proposed in (\ref{eqn:llrfunction_simo_non_coh_fad}),  we observe a quasi-linear complexity as a function of the input symbol size applied at the demodulator level, namely here $n = \{12, 24, 32, 48, 100\}$. Based on the average time complexity curves on Figure~\ref{fig:complexity_simo}, the complexity of the JED($M=4$) receiver is of the form
$\forall \epsilon \in(0,1), \exists \ a \in \mathbb{R}_{>0}$ tel que $\mathsf{T}(n)=a n^{1+\epsilon}=\mathcal{O}\left(n^{1+\epsilon}\right)$.
\vspace{-0.6em}

At $n=32$ for example, as in our simulations, the conventional receiver metric is around $39\times$ faster in terms of detection than the JED ($M=4$) metric over pure LOS condition ($\alpha=0$) in log domain versus $25\times$ against its max-log domain version. This makes the JED metric in max-log domain about $1.5 \times$ faster than the log-domain version.
It should also be emphasized that by also increasing $M$ (e.g., $M=8$), we'll expect the complexity to escalate, making the JED approach difficult to handle.


\subsection{Current Limitations and Future Work}
This proposed joint detection approach has inherent limitations. It becomes computationally prohibitive for higher-order modulation schemes like 16-QAM (where $|\boldsymbol{\chi}|=2^4$). With a joint detection block size of $M=4$, this would require processing $|\boldsymbol{\chi}|^M$ = 65,536 possible symbol vector combinations, which is computationally intractable. In such scenarios, reducing the detection block size (e.g., to $M=2$) becomes necessary to maintain feasible complexity.
\begin{figure*} [!ht]
    \centering
  \subfloat[\scriptsize {MIMO Rayleigh Block Fading channel, 5G Polar BICM. }\label{fig:mimo_polar_ray}]{
    \resizebox{0.40\linewidth}{!}{\pgfplotsset{
    tick label style={font=\footnotesize},
    label style={font=\footnotesize},
    legend style={nodes={scale=0.6, transform shape}, font=\footnotesize, legend cell align=left},
    every axis/.append style={line width=0.5pt}
}


\begin{tikzpicture}[spy using outlines=
      	{circle, magnification=3, connect spies}]
              \begin{semilogyaxis}[xlabel=\footnotesize{SNR (dB)},
            ,ylabel=\footnotesize{BLER},
            legend pos=north east,
            grid=both,
             minor tick num=3,
            width=6.8cm,
            height=5cm,
            ymax=0.7, ymin=0.001,xmin=5, xmax=25
    ]

  \addplot+[smooth,teal, line  width=0.8pt, mark options={scale=1, fill=teal, solid},mark=otimes]
    table[row sep=crcr]{%
    6	0.142857142857143\\
    10	0.0381970970206264\\
    16	0.0062\\
    24	0.001\\
  };
  \addlegendentry{Perfect CSI}

\addplot+[smooth, densely dotted, gray, line  width=0.8pt, mark options={ scale=1, fill=gray, solid},mark=otimes]   table[row sep=crcr]{%
6	0.380228136882129\\
10	0.107991360691145\\
16	0.0211014982063727\\
20	0.0075\\
24	0.0031\\
};
\addlegendentry{No CSI, Conv.}

\addplot+[smooth, dashed, violet,line  width=0.8pt, mark options={scale=1, fill=violet, solid},mark=otimes]
table[row sep=crcr]{%
6	0.234741784037559\\
10	0.0618046971569839\\
16	0.0103327133705311\\
18	0.0062\\
24	0.0014\\
};
\addlegendentry{No CSI, JED, $M=4$}

\end{semilogyaxis}

\end{tikzpicture}}
  }
\subfloat[\scriptsize {MIMO Rayleigh block Fading channel, 5G LDPC BICM.}\label{fig:mimo_LDPC_ray}]{
  \resizebox{0.40\linewidth}{!}{\pgfplotsset{
    tick label style={font=\footnotesize},
    label style={font=\footnotesize},
    legend style={nodes={scale=0.6, transform shape}, font=\footnotesize, legend cell align=left},
    every axis/.append style={line width=0.5pt}
}


\begin{tikzpicture}[spy using outlines=
      	{circle, magnification=3, connect spies}]
              \begin{semilogyaxis}[xlabel=\footnotesize{SNR (dB)},
            ,ylabel=\footnotesize{BLER},
            legend pos=north east,
           legend columns=1,
           legend pos=north east,
            grid=both,
             minor tick num=3,
            width=6.8cm,
            height=5cm,
            ymax=0.7, ymin=0.001,xmin=10, xmax=30
    ]

  \addplot+[smooth,teal, line  width=0.8pt, mark options={scale=1, fill=teal, solid},mark=otimes]
    table[row sep=crcr]{%
    11	0.0946969696969697\\
    13	0.0608272506082725\\
    19	0.0133226751931788\\
    25	0.0039\\
    29	0.0018\\
  };
  \addlegendentry{Perfect CSI}

\addplot+[smooth, densely dotted, gray, line  width=0.8pt, mark options={ scale=1, fill=gray, solid},mark=otimes]   table[row sep=crcr]{%
13	0.130890052356021\\
17	0.0526315789473684\\
23	0.0132152768600502\\
29	0.0037\\
};
\addlegendentry{No CSI, Conv.}

\addplot+[smooth, dashed, violet,line  width=0.8pt, mark options={scale=1, fill=violet, solid},mark=otimes]
table[row sep=crcr]{%
13	0.102774922918808\\
15	0.0542299349240781\\
25	0.0054\\
27	0.0035\\
29	0.0024\\
};
\addlegendentry{No CSI, JED, $M=4$}

%

%
%

%
%
%

\end{semilogyaxis}

\end{tikzpicture}}
  }
   \caption{Block Error Rate, 48 bits(TBs+CRC), NR POLAR BICM (CRC-aided successive-cancellation list decoder, List length=8), NR LDPC BICM (belief propagation decoder, iteration=30) QPSK modulation, 1 OFDM symbol, 4 PRBs, 48 REs (32 data, 16 DMRS), ($4\times4$) MIMO.}
   \label{fig:mimo_ray_block_fading_metric}
   \vspace{-6mm}
 \end{figure*}
\vspace{-0.4em}
\section{Conclusions}

This paper has introduced enhanced BICM receiver metrics for joint estimation–detection in short blocklength transmissions, specifically targeting scenarios with unknown channel state information and limited or sparse training resources. Through block-wise joint estimation–detection, we demonstrate significant improvements in performance and sensitivity compared to conventional receivers. Our analysis, conducted using full 5G transmitter and receiver chains with both Polar and LDPC coded transmissions under QPSK modulation, shows that even when reference signals are interleaved with coded data over a small number of OFDM symbols, precluding near-perfect channel estimation. Unlike conventional symbol-by-symbol detection in BICM systems, where the observation for a given coded bit is confined to the symbol in which it is conveyed,the proposed method performs block-wise joint detection over a sliding window of adjacent symbols to  fundamentally leverages their statistical dependencies; that is, the LLR for each coded bit incorporates information from multiple symbols rather than being confined to its host symbol. Performance evaluation in SIMO and SU-MIMO configurations confirms the effectiveness of this strategy in realistic base station receiver scenarios. Notably, detection windows spanning approximately four modulated symbols allow the proposed receivers to substantially outperform conventional ones, achieving detection performance approaching that of coherent receivers with perfect CSI.
Overall, the presented joint estimation–detection strategy built upon an enhanced likelihood metrics offers a significant advancement in receiver design for B5G/6G, especially in highly dynamic or resource-constrained environments.

\appendix
\label{appendix:llrfading}
\renewcommand{\thesection}{\Alph{section}.\arabic{subsection}}
\setcounter{section}{0}
\subsection{Metric derivation for non-coherent fading channel}
Assuming that $\theta_r$ is unknown and randomly distributed over

$[0, 2\pi)$, the conditional probability density function is defined as
\vspace{-0.4em}
\begin{align}
  \resizebox{0.42\textwidth}{!}{$
\begin{array}{r}
\mathrm q\left(\mathbf{x}, \mathbf{y}_r \right)
\coloneqq\displaystyle\frac{1}{2 \pi \operatorname{det} \Phi} \displaystyle\int_0^{2 \pi}
\exp \left(-\frac{1}{2}\left(\mathbf{y}_r-\ \mu\{\mathbf x, \theta_r\} \right)^{\dagger}\right. \\
\left.\Phi^{-1}\left(\mathbf{y}_r-\mu\{\mathbf x, \theta_r\}\right)\right) \mathrm{d} \theta_r \ .
\end{array}
$}
\end{align}
Saying  $\mu\{\mathbf x, \theta_r\}\coloneqq  \sqrt{\alpha} e^{j \theta_r}\mathbf x$, then
\begin{align}
\mathrm q\left(\mathbf{x}, \mathbf{y}_r \right)
=\displaystyle\frac{1}{2 \pi \operatorname{det} \Phi} \displaystyle\int_0^{2 \pi}
\exp \left(-\frac{1}{2}\left(\mathbf{y}_r-\sqrt{\alpha} e^{j \theta_r} \mathbf{x}\right)^{\dagger}\right. \\
\left.\Phi^{-1}\left(\mathbf{y}_r-\sqrt{\alpha} e^{j \theta_r} \mathbf{x}\right)\right) \mathrm{d} \theta_r  \ .\nonumber
\end{align}
\textbf{Covariance Matrix:}\\
Knowing that
  $\mathbf y_r - \sqrt{\alpha} e^{j \theta_r} \mathbf x = \sqrt{1-\alpha} \mathrm h_{i,f}\mathbf x + \mathbf z_i$,
then\\
\begin{equation}
\begin{aligned}
\Phi&=
\frac{1}{2}\mathbb{E}\left[\left(\sqrt{1-\alpha} \mathrm h_{r}^{(f)}\mathbf x + \mathbf z_i \right)\left(\sqrt{1-\alpha}\mathrm h_{r}^{(f)}\mathbf x + \mathbf z_i\right)^\dag\right],\\
&=  (1-\alpha)  \mathbf x\mathbf x^\dag \sigma^2_h + \sigma^2_z \mathbf I, \text{where } \sigma^2_h =1,\\
\end{aligned}
\end{equation}
\textbf{Determinant:}
\vspace{-0.3em}
\begin{equation}
\begin{aligned}\label{eqn:det1}
\operatorname{det} \Phi &= \operatorname{det}\left( (1-\alpha)  \mathbf x\mathbf x^\dag + \sigma^2_z \mathbf I \right),\\
 &= \operatorname{det}\left( \sigma^2_z \mathbf I + (1-\alpha)  \mathbf x\mathbf x^\dag \right),
\end{aligned}
\end{equation}
\begin{lemma} \cite[{\em Sylvester's determinant identity}]{Sylvester1851}.\\
when we have a matrix of the form: $\mathbf{A} = \mathbf{B} + \mathbf{u} \mathbf{v}^{\dagger}$, where $\mathbf{u}$ and $\mathbf{v}^\dagger$ are column vectors, we can apply the determinant formula for a matrix perturbed by a rank-1 term (i.e., the {\em rank-1 matrix} determinant formula or Sylvester's determinant Lemma), where $\mathbf{B}$ is an invertible square matrix of dimension $N \times N$, $\mathbf{u}$ and $\mathbf{v}^{\dagger}$ are complex column vectors of dimension $N \times 1$, and $\mathbf{u} \mathbf{v}^{\dagger}$ is thus a rank-1 matrix. The determinant of $\mathbf{A}$ is given by:
    \begin{equation} \label{eqn:det}
    \operatorname{det}(\mathbf{A})\coloneqq\operatorname{det}(\mathbf{B}) \cdot\left(1+\mathbf{v}^{\dagger} \mathbf{B}^{-1} \mathbf{u}\right).
    \end{equation}
\end{lemma}
Regarding the Sylvester's determinant Lemma, (\ref{eqn:det1}) then becomes,
\begin{equation}
\begin{aligned}
\operatorname{det} \Phi &= \operatorname{det}\left( \sigma^2_z \mathbf I + (1-\alpha) \mathbf x\mathbf x^\dag \right) =\sigma_z^{2 N}\left(1+\frac{ (1-\alpha)\mathbf{x}^{\dagger} \mathbf{x}}{\sigma^2_z}\right),
\\&=\sigma_z^{2 N}\left(1+\frac{ (1-\alpha)\|\mathbf{x}\|^2}{\sigma^2_z}\right),
 \\&=\left(\frac{\mathrm{N_0}}{2}\right)^N\left(1+\frac{(1-\alpha)\|\mathbf{x}\|^2}{\frac{\mathrm{N_0}}{2}}\right),
 \\&=\left(\frac{\mathrm{N_0}}{2}\right)^{N-1}\left(\frac{\mathrm{N_0}}{2}+{(1-\alpha)\|\mathbf{x}\|^2}\right),
 \\&=\frac{1}{2}\left(\frac{\mathrm{N_0}}{2}\right)^{N-1}\left(\mathrm{N_0}+{2(1-\alpha)\|\mathbf{x}\|^2}\right),
 \\&=c\left(\mathrm{N_0}+{2(1-\alpha)\|\mathbf{x}\|^2}\right), \text{ where  $c=\displaystyle\frac{1}{2}\left(\frac{\mathrm{~N}_0}{2}\right)^{N-1}$},
 \\&\propto \mathrm{N_0}+{2(1-\alpha)\|\mathbf{x}\|^2}.
\end{aligned}
\end{equation}
In what follows, we will pose $\mathbf L_{\mathsf x}=\mathrm{N_0} + 2(1-\alpha)\|\mathbf{x}\|^2.$
\textbf{Inverse of $\mathbf \Phi$ :}\\
The covariance matrix $\mathbf{\Phi}$ involves the addition of two matrices, which is amenable to consider the use of matrix inversion lemmas,  Sherman-Morrison-Woodbury formula, or simply the Woodbury Matrix identity.
 \begin{lemma} \cite[{\em The Woodbury Matrix identity}]{Woodbury1950}.
 \begin{equation}
 (\mathbf A+\mathbf U \mathbf C \mathbf V)^{-1}\coloneqq \mathbf A^{-1}-\mathbf A^{-1} \mathbf U\left(\mathbf C^{-1}+\mathbf{V A}^{-1} \mathbf U\right)^{-1} \mathbf {V A}^{-1},
 \end{equation}
 where $\mathbf A$, $\mathbf U$, $\mathbf C$, and $\mathbf V$ are matrices with  comfortable dimensions: $\mathbf A$ is a $n\times n$ matrix, $\mathbf C$ is a $k\times k$ matrix, $\mathbf U$ is a $n\times k$ matrix, and $\mathbf V$ is a $k\times n$ matrix.
  \end{lemma}
Note that,  here, we have a special case where $\mathbf V$, $\mathbf U$ are vectors, consequently $\operatorname{rank}\{\mathbf x^\dag \mathbf x\}=1$.
\begin{equation}
\begin{array}{ l l l l l}
 \text{Saying :} & \displaystyle \mathbf A= \sigma^2_z \mathbf I &
  \displaystyle \mathbf C =(1-\alpha)\mathbf I  &
  \displaystyle \mathbf U= \mathbf x &
  \displaystyle \mathbf V= \mathbf x^\dag  .
\end{array}
\end{equation}
\begin{equation}
\begin{aligned}
\Phi^{-1} &\coloneqq \left (\mathbf A+\mathbf U \mathbf C \mathbf V\right)^{-1},\\&
  =\frac{2}{\mathrm N_0}\mathbf I- \frac{2}{\mathrm N_0} \mathbf x
\left(\frac{2(1-\alpha)}{{\mathrm N_0} +2(1-\alpha)\left\|\mathbf x \right\|^2} \right)\mathbf x^\dag.
\end{aligned}
\end{equation}
Saying  $\beta_x = \frac{2(1-\alpha) }{\mathrm {N}_0(\mathrm {N}_0 +2(1-\alpha) \left\|\mathbf x \right\|^2)}$,  then $\Phi^{-1} =\frac{2}{\mathrm {N}_0}- 2 \mathbf x
\beta_x\mathbf x^\dag.$\\

\textbf{Likelihood function :}
\vspace{-1em}
\begin{equation}
\begin{aligned}
\mathrm q\left(\mathbf{x},\mathbf y_r\right)&=\displaystyle \frac{1}{2 \pi \operatorname{det} \Phi}
 \displaystyle \int_{0}^{2\pi}\exp \left(-\frac{1}{2}\left(\mathbf y_r- \sqrt{\alpha} e^{j \theta_r}\mathbf x\right)^\dag \left(\frac{2}{\mathrm{N_0}}\right.\right. \\&
\left.\left. - 2 \mathbf x
\mathbf \beta_x\mathbf x^\dag\right)\left(\mathbf y_r- \sqrt{\alpha} e^{j \theta_r}\mathbf x \right)\right)\mathrm{d} \theta_r,\\&
=\displaystyle \frac{1}{2 \pi \operatorname{det} \Phi}\displaystyle
 \int_{0}^{2\pi}\exp \left(-\frac{1}{\mathrm{N_0}}\left\|\mathbf y_r- \sqrt{\alpha} e^{j \theta_r}\mathbf x\right\|^2 \right. \\&
\left.+ \beta_x \left\|\left(\mathbf y^\dag_i- \sqrt{\alpha} e^{-j \theta_r}\mathbf x^\dag\right)\mathbf x\right\|^2\right)\mathrm{d} \theta_r.
\end{aligned}
\end{equation}
Extending the terms into the exponential, ignoring those that are independent of $\mathbf x$,
the likelihood function is equivalent to
\vspace{-1em}
\begin{equation}
\begin{array}{r}
\mathrm q\left(\mathbf{x},\mathbf y_r\right)
=\displaystyle \frac{1}{2 \pi \operatorname{det} \Phi}\exp \left(- \alpha \left\|\mathbf x \right\|^2 \left(\frac{1}{\mathrm {N}_0} - \beta_x \left\| \mathbf x\right\|^2\right) \right. \\
\left.
+ \beta_x \left| \mathbf x^\dag \mathbf y_r \right|^2\right) \displaystyle \int_{0}^{2\pi}\exp \left( 2\sqrt{\alpha} \left(\frac{1}{\mathrm {N}_0} - \beta_x \left\| \mathbf x\right\|^2\right)\right. \\
\left.|
  \mathbf x^\dag\mathbf y_r|\mathrm{cos}\left({\phi_i} + \theta_r \right)\right)\mathrm{d} \theta_r.
\end{array}
\end{equation}
knowing that $\displaystyle \frac{1}{\pi}\int_{{\varphi}=0}^{\pi}\exp(zcos(\varphi))\mathrm{d} \varphi=\operatorname{I_0(z)} $ \cite{Gradshteyn95}.
\begin{equation}
\begin{aligned}
&\mathrm q\left(\mathbf{x},\mathbf{y}_r\right)
=\frac{1}{2 \pi \operatorname{det} \Phi}\exp \left(- \alpha \left\|\mathbf x\right\|^2 \left(\frac{1}{\mathrm{N_0}} - \beta_x \left\| \mathbf x\right\|^2\right)\right. \\&
\left. + \beta_x \left| \mathbf x^\dag \mathbf y_r \right|^2\right)\times2\pi\times\operatorname{I_0}\left( 2\sqrt{\alpha} \left(\frac{1}{\mathrm{N_0}} - \beta_x \left\| \mathbf x\right\|^2\right)\left|\mathbf x^\dag\mathbf y_r\right|\right),\\&
= \frac{1}{\operatorname{det} \Phi}\exp \left(- \alpha \left\|\mathbf x\right\|^2 \left(\frac{1}{\mathrm{N_0}} - \beta_x \left\| \mathbf x\right\|^2\right) + \beta_x \left| \mathbf x^\dag \mathbf y_r \right|^2\right)\\&\times\operatorname{I_0}\left( 2\sqrt{\alpha} \left(\frac{1}{\mathrm{N_0}} - \beta_x \left\| \mathbf x\right\|^2\right)\left|\mathbf x^\dag\mathbf y_r\right|\right).
\end{aligned}
\end{equation}
Then after ignoring multiplicative term that are independent of $\mathbf x$, it comes  \begin{equation}\label{eqn:llrfunction_proof}
 \begin{aligned}
 &\mathrm q\left(\mathbf{x},\mathbf y_r\right)\propto \frac{1}{\mathbf L_\mathsf x}\exp \left(- \alpha \left\|\mathbf x \right\|^2\left(\frac{1}{\mathrm {N}_0} - \mathbf \beta_x \left\| \mathbf x\right\|^2\right) \right. \\& \left.  + \mathbf \beta_x \left| \mathbf x^\dag \mathbf y_r \right|^2\right)\times\operatorname{I_0}\left( 2\sqrt{\alpha} \left(\frac{1}{\mathrm {N}_0} - \mathbf \beta_x \left\| \mathbf x\right\|^2\right)\left|\mathbf x^\dag\mathbf y_r\right|\right).
 \end{aligned}
 \end{equation}
Expressing $\mathbf \beta_x$  w.r.t.  \ $\mathbf L_\mathsf{x}$, we have the relation
\begin{equation}
\begin{aligned}
\mathbf \beta_x&=
 \frac{2(1-\alpha) }{{\mathrm {N_0}}(\mathrm {N_0} +2(1-\alpha) \left\|\mathbf x \right\|^2)}
 = \frac{2(1-\alpha) }{{\mathrm N_0} \mathbf L_{\mathsf x}}
 =\frac{\mathbf L_{\mathsf x} - \mathrm {N_0}}{\left\|\mathbf x \right\|^2 \mathrm {N_0} \mathbf L_{\mathsf x}},\\&
 = \frac{\mathbf L_{\mathsf x}}{\left\|\mathbf x \right\|^2 \mathrm {N_0} \mathbf L_{\mathsf x}} - \frac{ \mathrm {N_0}}{\left\|\mathbf x \right\|^2\mathrm {N_0} \mathbf L_{\mathsf x}}
=\frac{1}{\left\|\mathbf x \right\|^2 \mathrm {N_0}} - \frac{ 1}{\left\|\mathbf x \right\|^2\mathbf L_{\mathsf x}}.
\end{aligned}
\end{equation}

\begin{equation}
\begin{aligned}
&\mathrm q\left(\mathbf{x},\mathbf{y}_r\right)\\=
&\frac{1}{\mathbf L_{\mathsf x}}\exp \left(- \alpha \left\|\mathbf x\right\|^2 \left(\frac{1}{\mathrm{N_0}} -(\frac{1}{\cancel{\left\|\mathbf x \right\|^2} \mathrm{N_0}} -\frac{ 1}{\cancel{\left\|\mathbf x \right\|^2}\mathbf L_{\mathsf x}})\cancel{\left\| \mathbf x\right\|^2}\right)  \right. \\& +\left. \beta_x \left| \mathbf x^\dag \mathbf y_r \right|^2\right)\operatorname{I_0}\left( 2\sqrt{\alpha} \left(\frac{1}{\mathrm{N_0}} - (\frac{1}{\cancel{\left\|\mathbf x \right\|^2} \mathrm{N_0}} \right.\right. \\& \left.\left. -\frac{ 1}{\cancel{\left\|\mathbf x \right\|^2}\mathbf L_{\mathsf x}})\cancel{\left\|\mathbf x \right\|^2}\right)\left|\mathbf x^\dag\mathbf y_r\right|\right).
\end{aligned}
\end{equation}
Thus the likelihood function is simply:
\begin{equation}\label{eqn:llrfunction_simplified}
\begin{aligned}
\mathrm q\left(\mathbf{x},\mathbf{y}_r\right)&= \frac{1}{\mathbf L_{\mathsf x}}\exp \left(  -\frac{\alpha \left\|\mathbf x\right\|^2 }{\mathbf L_{\mathsf x}}  + \mathbf \beta_x \left| \mathbf x^\dag \mathbf y_r\right|^2\right)\\&\times\operatorname{I_0}\left(  \frac{2\sqrt{\alpha}}{\mathbf L_{\mathsf x}}\left|\mathbf x^\dag\mathbf y_r\right|\right). ~\ \qedhere 
\end{aligned}
\end{equation}

\subsection{Finit-Blocklength Bounds}
Herein, we present the finite-blocklength information theory tools. An outer bound, derived from the metaconverse theorem (cf. \cite[Th. 28]{Polyanskiy2010}), is introduced, while an inner bound is established using the RCUs bound \cite[Th. 1]{Martinez2011}.
 \begin{theorem}\cite[RCU bound, Th. 1]{Martinez2011}.\\
 Let denote the random vectors via $\mathbf X = [\mathrm X_1, X_2, \ldots, \mathrm X_N]$, and their vector realizations via  $\mathbf x = [\mathrm x_1, \mathrm x_2, \ldots, \mathrm x_N]$. \\
To set the achievability bound, let define the generalized information density as
\begin{equation}
\imath_s\left(\mathbf{x}, \mathbf{y}\right)\coloneqq\ln \frac{\mathrm q\left(\mathbf{x}, \mathbf{y}\right)^s}{\mathbb{E}\left[\mathrm q\left(\mathbf{X}^\prime, \mathbf{y}\right)^s\right]}.
\end{equation}

In the domain of information theory, the Gallager exponent, represented by $s > 0$, characterizes a pivotal factor. The expectation relates to the random vector ${{\mathbf X}^\prime}$ having N-dimensional i.i.d. components. Over a memoryless channel, the decoding metric $\mathrm q\left(\mathbf{x}, \mathbf{y}\right) =\prod_{n=1}^N \mathrm q\left(\mathrm{x}_n, \mathrm{y}_n\right)$. The random coding unions (RCUs) posits that, for a specified rate $R$, the \texttt{upper bound} on the average error probability  is defined as:
\begin{equation}\label{eqn:rcu}
\epsilon \leq \inf _{s>0} \mathbb{E}\left[e^{-\left[\imath_s\left(\mathbf{x}, \mathbf{y}\right)-\ln \left(2^{R \times N}-1\right)\right]^{+}}\right],
\end{equation}
where $[u]^{+} \coloneqq \max (0, u)$.\\
The maximum likelihood decoding metric is shown to be
$\mathrm q(\mathbf{x}, \mathbf{y})=p_{\mathbf{Y} \mid \mathbf{X}, {\mathbf{H}}}\left(\mathbf{y}\mid \mathbf{x}, {\mathbf{h}}\right)$. Thus, the underlying decoding metric is expressed as $\mathrm q(\mathbf{x}, \mathbf{y}) \propto \exp \left(-\frac{1}{\mathrm {N}_0 }\lvert|\mathbf y-\mathbf h \mathbf x\rvert|^2\right)$.

\end{theorem}

 \begin{theorem}\cite[Metaconverse (Verdú–Han) bound, Th. 28]{Polyanskiy2010}.\\
To set the converse bound, consider :
\begin{equation}
\jmath_s\left(\mathbf{x}, \mathbf{y}\right)\coloneqq\ln \frac{\mathrm q\left(\mathbf{x}, \mathbf{y}\right)^s}{\mathbb{E}\left[\mathrm q\left(\mathbf{X}^\prime, \mathbf{y}\right)^s\right]^{1 / s}}=\frac{1}{s} \imath_s\left(\mathbf{x}, \mathbf{y}\right)
\end{equation}

Subsequently, for a given rate $R$, the \texttt{lower bound} on the average error probability is given as follows.
\begin{equation}
\epsilon \geq \sup _{s>0} \max_{\lambda} \mathrm{P}\left[ \jmath_s\left(\mathbf{x}, \mathbf{y}\right) \leq \lambda\right]-e^{\lambda-R\times N}.
\end{equation}


\end{theorem}

\ifCLASSOPTIONcaptionsoff
  \newpage
\fi

\balance






\vfill

\end{document}